\documentclass[american,notitlepage]{iopart}
\usepackage[T1]{fontenc}
\usepackage[latin9]{inputenc}
\usepackage{color}
\usepackage{babel}
\usepackage{prettyref}
\usepackage{amstext}
\usepackage{graphicx}
\usepackage{feyn}
\usepackage[unicode=true,pdfusetitle,
 bookmarks=true,bookmarksnumbered=false,bookmarksopen=false,
 breaklinks=false,pdfborder={0 0 1},backref=false,colorlinks=false]
 {hyperref}
\hypersetup{
 colorlinks=true,linkcolor=blue,citecolor=blue,urlcolor=blue,filecolor=blue}

\makeatletter

\providecommand{\tabularnewline}{\\}

\usepackage{iopams}
\usepackage{setstack}

\usepackage{stackrel}

\newrefformat{cap}{\hyperref[#1]{Figure~\ref{#1}}}
\newrefformat{fig}{\hyperref[#1]{Figure~\ref{#1}}}
\newrefformat{tab}{\hyperref[#1]{Table ~\ref{#1}}}
\newrefformat{sec}{\hyperref[#1]{Section~\ref{#1}}}
\newrefformat{sub}{\hyperref[#1]{Section~\ref{#1}}}
\newrefformat{cha}{\hyperref[#1]{Chapter~\ref{#1}}}
\newrefformat{app}{\hyperref[#1]{Appendix~\ref{#1}}}

\makeatletter
\newsavebox{\@brx}
\newcommand{\llangle}[1][]{\savebox{\@brx}{\(\m@th{#1\langle}\)}%
  \mathopen{\copy\@brx\kern-0.5\wd\@brx\usebox{\@brx}}}
\newcommand{\rrangle}[1][]{\savebox{\@brx}{\(\m@th{#1\rangle}\)}%
  \mathclose{\copy\@brx\kern-0.5\wd\@brx\usebox{\@brx}}}
\makeatother

\makeatother

\begin{document}
\global\long\def\T{\mathrm{T}}%
\global\long\def\D{\mathcal{D}}%
\global\long\def\o{\mathcal{O}}%
\global\long\def\N{\mathcal{N}}%
\global\long\def\Z{\mathcal{Z}}%
\global\long\def\E{\mathcal{E}}%
\global\long\def\C{\mathbb{C}}%
\global\long\def\R{\mathbb{R}}%
\global\long\def\tr{\mathrm{tr}}%

\title{Momentum-dependence in the infinitesimal Wilsonian renormalization
group}
\author{Moritz Helias}
\address{Institute of Neuroscience and Medicine (INM-6) and Institute for Advanced
Simulation (IAS-6) and JARA BRAIN Institute I, Jülich Research Centre,
Jülich, Germany}
\address{Department of Physics, Faculty 1, RWTH Aachen University, Aachen,
Germany}
\begin{abstract}
Wilson's original formulation of the renormalization group is perturbative
in nature. We here present an alternative derivation of the infinitesimal
momentum shell RG, akin to the Wegner and Houghton scheme, that is
a priori exact. We show that the momentum-dependence of vertices is
key to obtain a diagrammatic framework that has the same one-loop
structure as the vertex expansion of the Wetterich equation. Momentum
dependence leads to a delayed functional differential equation in
the cutoff parameter. Approximations are then made at two points:
truncation of the vertex expansion and approximating the functional
form of the momentum dependence by a momentum-scale expansion. We
exemplify the method on the scalar $\varphi^{4}$-theory, computing
analytically the Wilson-Fisher fixed point, its anomalous dimension
$\eta(d)$ and the critical exponent $\nu(d)$ non-perturbatively
in $d\in[3,4]$ dimensions. The results are in reasonable agreement
with the known values, despite the simplicity of the method.
\end{abstract}
\maketitle

\section{Introduction\label{sec:Introduction}}

The renormalization group (RG) is a standard tool to study phase transitions
in statistical physics as well as to investigate renormalizable field
theories in high energy physics. Its field-theoretical formulation
has initially been applied to renormalizable theories only \cite{GellMann1954};
those, in which all physical quantities can be expressed in terms
of a few renormalized parameters. One here exploits the arbitrariness
of the chosen scale at which the renormalized parameters are fixed
to derive a differential equation in this scale parameter \cite{Amit84,ZinnJustin96,Vasiliev98}.
It was realized later that the field-theoretical formulation is also
applicable to non-renormalizable theories \cite{Kazakov88_440}. Recent
examples of non-renormalizable models describe erosion of landscapes
\cite{Antonov17_193,Duclut17_012149}. These models are structurally
similar to a variant of the KPZ equation \cite{Kardar84}, the model
by \cite{Pavlik94_303}, whose RG flow was first shown by Antonov
and Vasiliev to require arbitrary many couplings \cite{Antonov95_485}.

Conceptually quite different is the idea of Wilson's renormalization
group, which studies how the action that describes a theory changes
as short-ranged degrees of freedom are marginalized out. The interpretation
is thus more intuitive. Initially, this scheme has been presented
as an iterative method, where in each step a fixed fraction of the
the momentum space is integrated out \cite{Wilson74_75,Wilson75_773}.
A hard cutoff separates the marginalized, short-distance modes from
the long-distance modes that remain after the marginalization step.
One studies how the action evolves as the cutoff is lowered. This
approach has the merit of a clear interpretation as a systematic coarse-graining.
Also, the procedure is applicable to theories that originally possess
a high momentum cutoff. Those appear, for example, in condensed matter
problems, where the lattice spacing limits the magnitude of all momenta.
The Wilsonian RG, too, is applicable also to non-renormalizable theories.

The classical computation in the Wilsonian RG, however, relies on
a diagrammatic perturbation expansion. For this reason it is tied
to the computation of properties of systems close to their upper critical
dimension $d_{c}$, the dimensionality above which non-Gaussian terms
can be neglected. In the vicinity below $d_{c}$, interaction vertices
are small, presenting a natural expansion parameter. The $\epsilon$-expansion,
operating in $d_{c}-\epsilon$ dimensions, exploits this fact. Quantities
of interest must ultimately be computed from typically divergent series
in $\epsilon$, requiring appropriate resummation. Also, computations
become complicated beyond one loop order due to the presence of the
cutoff. In practice, one therefore often resorts to the field-theoretical
formulation, even in the context of condensed matter problems.

Early on, Wegner and Houghton \cite{Wegner73_401} presented a continuous
version of the Wilsonian RG by employing a sharp cutoff and by marginalizing
only over an infinitesimally thin momentum shell. But the simplicity
of the diagrammatic formulation by Wilson was lost to some extent
due to the employed projector formalism. The authors noted difficulties
of a sharp cutoff: it generates a non-analytical momentum dependence
of the vertices at vanishing momenta, which correspond to long-range
interactions \cite[p. 153]{Wilson74_75}. This general problem of
treating the momentum dependence within the Wilsonian RG has been
noted early on \cite[cf. Fig 11.1 and surrounding text]{Wilson74_75}.
A sharp cutoff is, however, needed ensure that the action maintains
its rescaling invariance, the independence of physical results under
the transform $\varphi\to z\,\varphi$ \cite{Morris96_477}. This
property must be kept in order to compute the anomalous dimension.

Morris \cite{Morris96_477} showed that there is in fact no fundamental
problem of using a hard cutoff, if combined with the flow equation
for the effective action \cite{WETTERICH93_90}, the Legendre transformed
Helmholtz free energy. A momentum-scale expansion can be made for
the flow equation, where the momentum scale $|k|$ is used as the
expansion parameter. However, the author pointed out a conceptual
problem: despite being an expansion for small external momenta, vertices
evaluated at momenta as large as the cutoff are required to close
the equations. Also, critical exponents turned out to be not as accurate
as obtained with methods of comparable complexity. Kopietz \cite{Kopietz2001_493}
showed that the exact Wetterich flow equation for the effective action
\cite{WETTERICH93_90} indeed can be used to systematically derive
results beyond one-loop order. A drawback of this method is the requirement
to compute diagrams with more than a single loop.

The generality of the Wilsonian RG, working for renormalizable as
well as for non-renormalizable theories, as well as its intuitive
interpretation in terms of coarse-graining and its efficient diagrammatic
formulation are desirable features. However, it has been pointed out
that a momentum-scale expansion was incompatible with the closely
related Wegner and Houghton RG equation, because the latter contained
tree-diagrams besides one-particle irreducible components. The reason
is that tree diagrams vanish only for zero external momenta, as the
momentum on the connecting propagator is constrained to reside above
the cutoff.

The main motivation of this work is three-fold: First, to clarify
where approximations are made when studying the a-priori exact formulation
of the infinitesimal Wilsonian momentum-shell RG. Second, to derive
a tower of diagrammatic expressions that is logically consistent with
the idea of a systematic coarse-graining and a hard cutoff between
marginalized and non-marginalized modes. Third, to recheck if a momentum-scale
expansion is compatible with this framework. The latter question arises
because it seems unintuitive that there was a qualitative difference
between the Wegner-Houghton and the Wetterich equation with a hard
cutoff, as sometimes stated in the literature. 

To this end we show that in fact a continuous Wilsonian RG equation
with sharp cutoff can be derived naturally without the projector formalism
by Wegner and Houghton and be interpreted diagrammatically. Our derivation
in particular shows that there is no approximation implied by going
to only a single loop. We find that again tree diagrams appear, but
their treatment is actually quite simple: one can systematically keep
only those tree diagrams that are needed to integrate the remainder
of the flow equation that is composed of individual vertices with
a single pair of legs contracted. This leads to a delayed functional
differential equation in the cutoff parameter, where the delays depend
on the external momenta. The structure of the resulting diagrams is
in fact identical to those of the Wetterich equation. Tree diagrams
disappear from the final effective long-range theory, as one evaluates
all quantities on the long-distance scale of small momenta.

To perform actual computations, however, one is still forced to apply
approximations, similarly as in the case of the Wetterich equation.
We here employ two approximations to arrive at closed-form results:
First, a truncation in the power of the field, only keeping a limited
number of vertices for which we compute the flow. This approximation
is naturally guided by the relevance of each term as indicated by
its engineering dimension. Second, the momentum-dependence of the
vertices is approximated in a momentum-scale approximation, as introduced
by Morris \cite{Morris96_477} and employed in the context of the
functional RG \cite{Hasselmann04_101103,Ledowski04_101103}.

We exemplify this method by computing an approximation of the critical
exponent $\nu$ and the anomalous dimension $\eta$ of the scalar
$\varphi^{4}$-theory. The computation requires only elementary spherically
symmetric and uniaxial one-loop integrals. We show that the flow equations
can be evaluated far off the upper critical dimension $d_{c}=4$,
directly obtaining results for any dimension $d\in[3,4]$. This demonstrates
the non-perturbative nature of the equation. We find that the approximation
for the critical exponent $\eta$ is in between the values for the
$\epsilon$-expansion of orders $\epsilon^{2}$ and $\epsilon^{3}$
and slightly better than the result obtained by Morris \cite{Morris96_477}.
It is, moreover, far better than the second order $\epsilon$-expansion,
despite only requiring one-loop integrals. Also there is no divergent
series to be resumed to obtain estimates for critical exponents.

The results in \prettyref{sec:Results} are presented in a self-contained
manner. The initial sections set up the notation and present a coherent
exposure of the basic idea of the renormalization group: In \prettyref{sec:Model-system}
we introduce the notation for the field theory to be considered and
the $\varphi^{4}$-theory as a particular example. \prettyref{sec:Definition-of-coarse-graining}
defines the procedure of coarse-graining in Fourier domain. \prettyref{sec:The-exact-flow}
derives an exact flow equation for the coarse-grained action. \prettyref{sec:Relation-to-Wetterich}
motivates the comparison between the Wetterich and the Wegner-Houghton
scheme. \prettyref{sec:Two-classes-of-graphs} discusses the two distinct
classes of graphs, single loops and trees, that contribute and shows
how they are combined to obtain one-loop diagrams of identical structure
as those of the Wetterich equation. It shows how the momentum dependence
of effective vertices couples different decimation steps in the flow
equation. \prettyref{sec:Decimating-an-infinitesimal-shell} derives
concrete expressions for these diagrams. Rescaling of momenta and
fields, required to obtain fixed points, is the topic of \prettyref{sec:Rescaling-of-momenta}.

The method is then illustrated on the example of the $\varphi^{4}$-theory,
obtaining a general expression for the momentum dependence of the
four-point vertex in \prettyref{sec:Momentum-dependence}, which is
reduced to a set of coupled differential equations for the interaction
in \prettyref{sec:Flow-of-the-momentum-dependent-interaction} and
for the self-energy in \prettyref{sec:Flow-of-the-self-energy}. \prettyref{sec:Fixed-points-and-critical-exponents}
determines the fixed points in dimensions $d\in[3,4]$ and computes
the critical exponents $\nu(d)$ and $\eta(d)$. \prettyref{sec:Summary-and-discussion}
summarizes the results in the light of the literature and compares
this interpretation of the Wilsonian RG to the $\epsilon$-expansion
\cite{Wilson74_75,Wilson75_773} and to the work by Wegner \& Houghton
\cite{Wegner73_401}.

\section{Results\label{sec:Results}}

We here chose a notation that should be easy to transfer to other
problems. For concreteness, we here choose the language of classical
statistical field theory, in particular to bosonic fields. In the
following \prettyref{sec:Model-system} we set up the language and
define elementary quantities.

\subsection{Model system\label{sec:Model-system}}

We assume a form of the action

\begin{eqnarray}
S[\varphi] & = & -\frac{1}{2}\,\varphi^{\T}G^{-1}\varphi,\label{eq:action_renorm_general}\\
 &  & +\sum_{n=3}^{\infty}\frac{1}{n!}\,\int_{k_{1}}\cdots\int_{k_{n}}\,S^{(n)}(k_{1},\ldots,k_{n})\,\varphi(k_{1})\cdots\varphi(k_{n}),\nonumber 
\end{eqnarray}
of a translation-invariant system, so that the quadratic part $G^{-1}\equiv-S^{(2)}$
and the bare interaction vertices $S^{(n)}$ all conserve momenta
\begin{eqnarray}
G^{-1}(k_{1},k_{2}) & \propto & \delta(k_{1}+k_{2}),\label{eq:momentum_conservation}\\
S^{(n)}(k_{1},\ldots,k_{n}) & \propto & \delta(k_{1}+\ldots+k_{n}).\nonumber 
\end{eqnarray}
As a particular example, we study the $\varphi^{4}$-theory

\begin{eqnarray}
S[\varphi]= & - & \frac{1}{2}\int_{k}\,\varphi(-k)\,(r^{(0)}+r^{(2)}\,k^{2})\,\varphi(k)\label{eq:phi4_theory}\\
 & - & u^{(0)}\,\int_{k_{1}}\int_{k_{2}}\int_{k_{3}}\,\varphi(k_{1})\,\varphi(k_{2})\,\varphi(k_{3})\,\varphi(-k_{1}-k_{2}-k_{3}),\nonumber 
\end{eqnarray}
as a prototypical system, for which the Gaussian part $G^{-1}$ is
given by $r^{(0)}+r^{(2)}\,k^{2}$ and there is only a single bare
interaction vertex $u^{(0)}\,\delta(k_{1}+\ldots+k_{4})$. The superscripts
here refer to the order in $k$ of the momentum dependence. A configuration
of the field $\varphi(k)$ is realized with probability of Boltzmann
form
\begin{eqnarray*}
p[\varphi] & \propto & \exp\big(S[\varphi]\big),
\end{eqnarray*}
and the system is described by the partition function
\begin{eqnarray}
\Z & = & \int D\varphi\,\exp\big(S[\varphi]\big).\label{eq:partition_function}
\end{eqnarray}

\subsection{Definition of coarse-graining\label{sec:Definition-of-coarse-graining}}

We assume the system lives on a space $r\in\mathbb{R}^{d}$ and the
elementary entities (e.g. spins) have a lattice spacing $a$. Correspondingly,
we have a high-momentum cutoff $\Lambda=\frac{\pi}{a}$, so that $|k|<\Lambda$.

We use the notation $\varphi_{<}$ for the coarse-grained field that
is defined in terms of the long-ranged degrees of freedom

\begin{eqnarray}
\varphi_{<}(r) & := & \int_{0\le|q|<\Lambda_{\ell}}\,\varphi(q)\,e^{iqr}\,,\label{eq:long_ranged_degf}
\end{eqnarray}
where we only take the Fourier integral up to the new cutoff $\Lambda_{\ell}:=\Lambda/\ell$
with an $\ell>1$. Here and in the following we use the short hand
$\int_{0\le|q|<\Lambda\ell^{-1}}=\int_{0\le|q|<\Lambda_{\ell}}\,\frac{d^{d}q}{(2\pi)^{d}}$.
Analogously we define the short-ranged degrees of freedom as 
\begin{eqnarray*}
\varphi_{>}(r) & := & \int_{\Lambda_{\ell}<|k|<\Lambda}\,\varphi(k)\,e^{ikr}\,.
\end{eqnarray*}
Here and in the following we will denote momenta as $q$ if they are
long-ranged, $0<|q|<\Lambda_{\ell}$, and those as $k$ which short
ranged, $\Lambda_{\ell}<|k|<\Lambda$.

The Fourier transform decomposes the fluctuations into long-ranged
$\varphi_{<}$ and short-ranged $\varphi_{>}$ ones; with the linearity
of the Fourier transform and the orthogonality of Fourier modes, it
yields a decomposition of the entire field into the direct sum
\begin{eqnarray*}
\varphi & = & \varphi_{<}+\varphi_{>}.
\end{eqnarray*}
We may ask which action $S_{\ell}$ effectively controls the coarse-grained
modes $\varphi_{<}$. This question reduces to a marginalization of
the fast modes and we define the action for the slow modes as
\begin{eqnarray}
\exp(S_{\ell}[\varphi_{<}]) & := & \int\,\D\varphi_{>}\,\exp(S[\varphi_{<}+\varphi_{>}]),\label{eq:coarse_grained_action}
\end{eqnarray}
where the notation $\int\,\D\varphi_{>}$ should be read as the integral
of all Fourier modes with $\Lambda_{\ell}<|k|<\Lambda$. We call this
step decimation. By this definition and the orthogonality of Fourier
modes we have
\begin{eqnarray*}
\int\D\varphi & = & \int\D\varphi_{<}\,\int\D\varphi_{>}
\end{eqnarray*}
so that the partition function \prettyref{eq:partition_function}
can be written in terms of the coarse-grained action alone
\begin{eqnarray*}
\Z & = & \int\D\varphi_{<}\,\exp(S_{\ell}[\varphi_{<}]).
\end{eqnarray*}
Performing the marginalization over a finite momentum shell, the modes
with Fourier coefficients $|k|\in[\Lambda_{\ell},\Lambda]$ are integrated
out. After this step, the range of momenta is limited to $0<|q|<\Lambda_{\ell}$.

To find a fixed point we need to bring the action into a comparable
form as the initial action, thus rescaling the length scale by $\ell$
and the momentum scale correspondingly as
\begin{eqnarray}
\ell q & =: & k_{\ell},\label{eq:def_rescaled_k}
\end{eqnarray}
so that there are two consecutive steps

\begin{eqnarray}
\label{eq:decimation_and_rescaling}\\
 & \begin{array}{c}
S[\varphi(k)]\\
|k|\in[0,\Lambda]
\end{array} & \stackrel{\text{decimation}}{\rightarrow}\begin{array}{c}
S_{\ell}[\varphi_{<}(q)]\\
|q|\in[0,\Lambda_{\ell}]
\end{array}\stackrel{\text{rescaling}}{\rightarrow}\begin{array}{c}
S_{\ell}[\varphi_{\ell}(k_{\ell})]\\
|k_{\ell}|=|\ell\,q|\in[0,\Lambda]
\end{array}.\nonumber 
\end{eqnarray}

\subsection{The exact flow equation\label{sec:The-exact-flow}}

The aim of this section is to derive a differential form for the Wilson
RG. To this end, we consider the difference between the coarse-grained
action $S_{\ell\cdot(1+\delta)}$, defined as \prettyref{eq:coarse_grained_action},
and the action $S_{\ell}$; the cutoff $\Lambda_{\ell}$ is thus lowered
by an infinitesimal amount controlled by $\delta>0$. The assumption
is thus that all modes $|k|>\Lambda_{\ell}$ have been marginalized
and we integrate over the next lower interval of modes
\begin{eqnarray}
\Lambda_{\ell\cdot(1+\delta)}\equiv\frac{\Lambda_{\ell}}{1+\delta} & < & |k|<\Lambda_{\ell}.\label{eq:def_shell-1}
\end{eqnarray}
To obtain a flow equation, consider the difference between the coarse-grained
action \prettyref{eq:coarse_grained_action} after this marginalization
step and before
\begin{eqnarray}
S_{\ell\cdot(1+\delta)}[\varphi_{<}]-S_{\ell}[\varphi_{<}] & = & \ln\,\Big(\int\,\D\varphi_{>}\,\exp\big(S_{\ell}[\varphi_{<}+\varphi_{>}]\big)\Big)\label{eq:single_step}\\
 &  & -S_{\ell}[\varphi_{<}]\nonumber \\
 & = & \ln\,\int\,\D\varphi_{>}\,\exp\big(S_{\ell}[\varphi_{<}+\varphi_{>}]-S_{\ell}[\varphi_{<}]\big).\nonumber 
\end{eqnarray}
The logarithm appearing on the right hand side, by the linked cluster
theorem (see e.g. \cite{ZinnJustin96} or \cite[Appendix A.3]{Kuehn18_375004}),
ensures that only connected diagrams need to be taken into account.
The integrand can be expanded as
\begin{eqnarray}
S_{\ell}[\varphi_{<}+\varphi_{>}]-S_{\ell}[\varphi_{<}] & = & \varphi_{>}^{\T}\,S_{\ell}^{(1)}[\varphi_{<}]+\frac{1}{2}\,\varphi_{>}^{\T}\,S_{\ell}^{(2)}[\varphi_{<}]\:\varphi_{>}\label{eq:expanded_integrand}\\
 &  & +\o(\varphi_{>}^{3}),\nonumber 
\end{eqnarray}
where $S_{\ell,k}^{(1)}\equiv\frac{\delta S_{\ell}}{\delta\varphi(k)}$
and $S_{\ell,-k,k}^{(2)}=\frac{\delta^{2}S_{\ell}}{\delta\varphi(-k)\delta\varphi(k)}$
denote the first and second functional derivatives of the action.
The first line in \prettyref{eq:expanded_integrand} leads to a Gaussian
integral in \prettyref{eq:single_step}. Now, each integration over
$\varphi_{>}$ has a momentum constrained to the shell $\Lambda_{\ell\cdot(1+\delta)}<|k|<\Lambda_{\ell}$
and thus contributes in proportion to the integrated volume in $k$-space,
which is $\propto\delta$, the thickness of the momentum shell.

Taking the limit $\delta\to0$ and considering only the infinitesimal
change of the action on the left hand side of \prettyref{eq:single_step}
as 
\begin{eqnarray}
\lim_{\delta\to0}\frac{1}{\delta}\big(S_{\ell\cdot(1+\delta)}-S_{\ell}\big) & \equiv & \ell\frac{dS_{\ell}}{d\ell},\label{eq:limit_delta_0}
\end{eqnarray}
all diagrams generated by the terms $\o(\varphi_{>}^{3})$ are $\o(\delta^{2})$,
because they need at least two propagators and thus two independent
momentum integrals to contract all $\varphi_{>}$ fields; these terms
therefore vanish in the considered limit. As a result, only the Gaussian
integral involving the first line of \prettyref{eq:expanded_integrand}
remains, leading to
\begin{eqnarray}
\ell\frac{dS_{\ell}}{d\ell}[\varphi_{<}] & = & \frac{1}{2}\,\int_{|k|=\Lambda_{\ell}}\,\ln\big(G_{\ell,-k,k}[\varphi_{<}]\big)\label{eq:exact_flow_eq}\\
 & + & \frac{1}{2}\,\int_{|k|=\Lambda_{\ell}}S_{\ell,-k}^{(1)}[\varphi_{<}]\,G_{\ell,-k,k}[\varphi_{<}]\,S_{\ell,k}^{(1)}[\varphi_{<}],\nonumber 
\end{eqnarray}
where we defined the propagator $G_{\ell,-k,k}[\varphi_{<}]:=\big(-S_{\ell,-k,k}^{(2)}[\varphi_{<}]\big)^{-1}$.
The first line results from the normalization term of the Gaussian
integral $-\frac{1}{2}\,\ln\,\prod_{|k|=\Lambda_{\ell}}\Big(-S_{\ell,-k,k}^{(2)}[\varphi_{<}]\Big)$.
This equation is an exact flow equation for the coarse-grained action
$S_{\ell}$. Apart from a different derivation and notation, this
equation is identical to the Wegner-Houghton equation \cite[their eq. (2.14)]{Wegner73_401}.

\subsection{Relation to the Wetterich equation\label{sec:Relation-to-Wetterich}}

What is the relation to the Wetterich equation (\cite{WETTERICH93_90},
for a review see \cite[eq. 2.19]{Berges02_223})? The latter being
an exact flow equation that is of one-loop structure. Both are functional
differential equations: The Wetterich equation for the effective action
$\Gamma$, the first Legendre transform of the free energy $\ln\Z$;
the Wegner-Houghton equation \prettyref{eq:exact_flow_eq} for the
Wilson effective action $S_{\ell}$ defined by \prettyref{eq:coarse_grained_action}.
In particular it may, at first sight, be surprising why eq. \prettyref{eq:exact_flow_eq}
contains two terms, while in the Wetterich equation only a single
term appears that produces diagrams with a single closed loop. This
apparent difference in the structure of the equations can be resolved
by considering the dependence on external momenta of the vertices
$S_{\ell}^{(n)}$, as shown in the following.

\subsection{Two classes of graphs: single loops and trees\label{sec:Two-classes-of-graphs}}

Expanding the action into vertices \prettyref{eq:action_renorm_general},
written for short as $S_{\ell}[\varphi]=\sum_{n}S_{\ell}^{(n)}\varphi^{n}/n!$,
the exact flow equation \prettyref{eq:exact_flow_eq} has two contributions: 

\paragraph*{Single loops }

The first line of \prettyref{eq:exact_flow_eq} has a one-loop structure
that becomes apparent when considering functional derivatives by $\varphi$;
the first derivative by $\varphi(q)$, for example, yields a tadpole
contribution
\begin{eqnarray*}
\frac{1}{2}\,\int_{|k|=\Lambda_{\ell}}S_{\ell,-k,q,k}^{(3)}G_{\ell,k,-k} & \equiv & 3\cdot\vertexlabel_{q}\;\Diagram{gf0flflu}
\qquad,\\
\end{eqnarray*}
which looks like a usual one-loop correction to the equation of state.
Taking further derivatives by $\varphi$, the first line of \prettyref{eq:exact_flow_eq}
produces the usual structure of one-loop diagrams. Considering the
second functional derivative by $\varphi(q)$ and $\varphi(-q)$,
with $\delta/\delta\varphi(-q)\,G_{\ell,k,-k}=G_{\ell,k,-k}S_{\ell,-k,-q,k}^{(3)}G_{\ell,k,-k}$,
one obtains
\begin{eqnarray}
 &  & \frac{1}{2}\,\int_{|k|=\Lambda_{\ell}}S_{\ell,-k,q,-q,k}^{(4)}G_{\ell,k,-k}\label{eq:self-energy-WH}\\
 &  & +\frac{1}{2}\,\int_{|k|=\Lambda_{\ell}}S_{\ell,-k,q,k}^{(3)}G_{\ell,k,-k}\,S_{\ell,-k,-q,k}^{(3)}G_{\ell,k,-k}.\nonumber 
\end{eqnarray}
Compared to the usual loopwise expansion \cite{ZinnJustin96,Vasiliev98}
(see \cite{Helias19_10416} Chapter XIV for a presentation in the
same notation), there are, however, three differences:
\begin{itemize}
\item the propagator $G_{\ell}\equiv(-S_{\ell}^{(2)})^{-1}$ and the vertices
$S_{\ell}^{(n)}$, being solutions of the differential equation \prettyref{eq:exact_flow_eq},
contain all fluctuation corrections from the scales $\Lambda_{\ell}<|k|<\Lambda$
that have already been integrated out;
\item the momentum integration is constrained to a single shell $\int_{|k|=\Lambda_{\ell}}$;
\item all propagator lines carry exactly the same momentum $k$.
\end{itemize}
These points hold in general also for higher derivatives. The last
point has implications for the dependence on external momenta: the
second line of \prettyref{eq:self-energy-WH} contributes only at
vanishing external momentum $q=0$ by the assumption of a theory with
momentum-conserving vertices $S_{-k,-q,k}^{(3)}\propto\delta(q)$
(cf. \prettyref{eq:momentum_conservation}). The first line contributes
to potentially all external momenta $q$.

The structure of the first diagram is a single vertex with two of
its legs contracted by a propagator. Such one loop diagrams have a
non-trivial flow equation, in the sense that multiple decimation steps
$\ell$ contribute to a vertex at a given and fixed set of external
momenta. To see this explicitly, we take as an example a portion of
the one-loop integral

\begin{eqnarray}
\nonumber \\
\Diagram{\vertexlabel^{\varphi_{<}(q_{1})} &  & \vertexlabel^{\quad\quad q_{1}+q_{2}+k}\\
gd & !f{}u\\
gu & !f{}d\\
\vertexlabel_{\varphi_{<}(q_{2})} &  & \vertexlabel_{\quad-k}
}
 & ,\label{eq:example_vertex_closed_loop-1}\\
\nonumber 
\end{eqnarray}
where the wiggly lines denote amputated legs and the full lines the
connection points of a propagator. Now consider the point $q_{1}=-q_{2}$
in momentum space that obeys momentum conservation of the amputated
legs. The full lines can then be closed by a propagator $G_{\ell}(k)$.
By momentum conservation at the vertex, any value of $k$ is allowed.
So all decimation steps $\ell$, where a momentum shell $|k|=\Lambda_{\ell}$
is marginalized, will contribute to the flow of the vertex function
at this set of external momenta.

\paragraph*{Trees}

The second line of \prettyref{eq:exact_flow_eq} comprises two components
$S_{\ell^{\prime}}^{(1)}$ that are connected by a single propagator
$G_{\ell^{\prime}}$, which we call reducible diagrams. These diagrams
therefore have tree topology. Expanded in vertices $S_{\ell^{\prime}}^{(n)}$,
this contribution is the sum of pairs of vertices connected by a single
line. The first derivative $\delta/\delta\varphi_{>}$ in $S_{\ell^{\prime}}^{(1)}$
produces the usual combinatorial factor $n$ for each term $S_{\ell}^{(n)}\varphi_{>}^{n}\,/\,n!$
of the expansion. The flow equation of these tree diagrams is trivial
in the sense that only a single decimation step contributes to each
tree: The decimation step $\ell^{\prime}$ of the momentum shell that
contains the momentum $k^{\prime}$ of the connecting propagator $G_{\ell^{\prime}}(k^{\prime})$,
which is $|k^{\prime}|=\Lambda_{\ell^{\prime}}$.

As an example consider the tree diagram
\begin{eqnarray}
 &  & \frac{1}{2}\,\int_{k,q}\,\frac{\varphi(-k)\,\varphi(-q)}{2!}\,S_{\ell^{\prime},-k,-q,k^{\prime}}^{(3)}\,G_{\ell^{\prime},k^{\prime},-k^{\prime}}\,S_{\ell^{\prime},-k^{\prime},q,k}^{(3)}\frac{\varphi(k)\,\varphi(q)}{2!}\label{eq:tree_diagram}\\
 & = & \Diagram{\vertexlabel^{-q}\quad & gd & f & gu\: & \vertexlabel^{q}\\
\vertexlabel^{-k}\quad & gu &  & gd\: & \vertexlabel^{k}
}
\quad.\nonumber 
\end{eqnarray}
The resulting contribution is an effective four-point vertex $S_{\ell^{\prime}}^{(4)}$
with external momenta $-q,-k,q,k$. The external momenta $(k,q)$
uniquely determine the momentum $k^{\prime}=k+q$ of the connecting
propagator and thus the decimation step $\ell^{\prime}$ by $|k^{\prime}|=\Lambda_{\ell^{\prime}}\equiv\Lambda/\ell^{\prime}$.
For the diagram to exist in the effective action $S_{\ell}$, the
corresponding shell must have been integrated, so $\ell\ge\ell^{\prime}$.
In summary,
\begin{itemize}
\item contributions from tree diagrams to the coarse-grained action can
be integrated trivially; they have a non-vanishing value whenever
the momentum $k^{\prime}$ of the propagator that connects the pair
of vertices of the tree is not below the current cutoff $|k^{\prime}|\ge\Lambda_{\ell}$;
\item as a consequence, the decimation step $\ell^{\prime}$ within which
a tree diagram is created depends on the momentum $k^{\prime}$ as
$|k^{\prime}|=\Lambda_{\ell^{\prime}}\equiv\Lambda/\ell^{\prime}$.
\end{itemize}

\paragraph*{Combining loops and trees}

Because only a single decimation step contributes to each tree, when
integrating the flow equation \prettyref{eq:exact_flow_eq}, tree
diagrams can directly be substituted in place of the effective vertices.
This allows the construction of composed diagrams, illustrated in
\prettyref{fig:Decimation-causes-flow-momentum}: In a first decimation
step at scale $\ell^{\prime}$, a tree diagram \prettyref{eq:tree_diagram}
is produced, creating an effective four point vertex $S_{\ell^{\prime}}^{(4)}$.
At scale $\ell>\ell^{\prime}$, the first line of \prettyref{eq:self-energy-WH}
contracts its two legs with momenta $k$ and $-k$ with the propagator
$G_{\ell,k,-k}$ to form a single loop, resulting in
\begin{eqnarray}
\label{eq:self-energy-WH-trees-inserted-1}\\
 &  & \frac{1}{2}\,\int_{|k^{\prime}|>\Lambda_{\ell}}\,\int_{|k|=\Lambda_{\ell}}S_{\ell^{\prime}(k^{\prime}),-k^{\prime},q,k}^{(3)}\,G_{\ell,k,-k}\,S_{\ell^{\prime}(k^{\prime}),-k,-q,k^{\prime}}^{(3)}\,G_{\ell^{\prime}(k^{\prime}),k^{\prime},-k^{\prime}}.\nonumber 
\end{eqnarray}
The structure and combinatorial factor is hence the same as the second
line of \prettyref{eq:self-energy-WH}, only that it is non-vanishing
also for $q\neq0$. The resulting diagrams in general have the following
properties:
\begin{itemize}
\item they have the one-loop structure known from the loopwise fluctuation
expansion;
\item at decimation step $\ell$ there is precisely one single-scale propagator,
a propagator whose momentum is constrained to the shell of the current
decimation step, $G_{\ell}(k)$ with $|k|=\Lambda_{\ell}$;
\item all other propagators $G_{\ell^{\prime}}$ belong to scales $\ell^{\prime}<\ell$
that have already been integrated out, so they carry momenta $|k^{\prime}|>\Lambda_{\ell}$;
\item combined with the contribution akin to the second line of \prettyref{eq:self-energy-WH},
the last statement extends to $\ell^{\prime}=\ell$.
\end{itemize}
\begin{figure}
\begin{centering}
\includegraphics[width=0.9\textwidth]{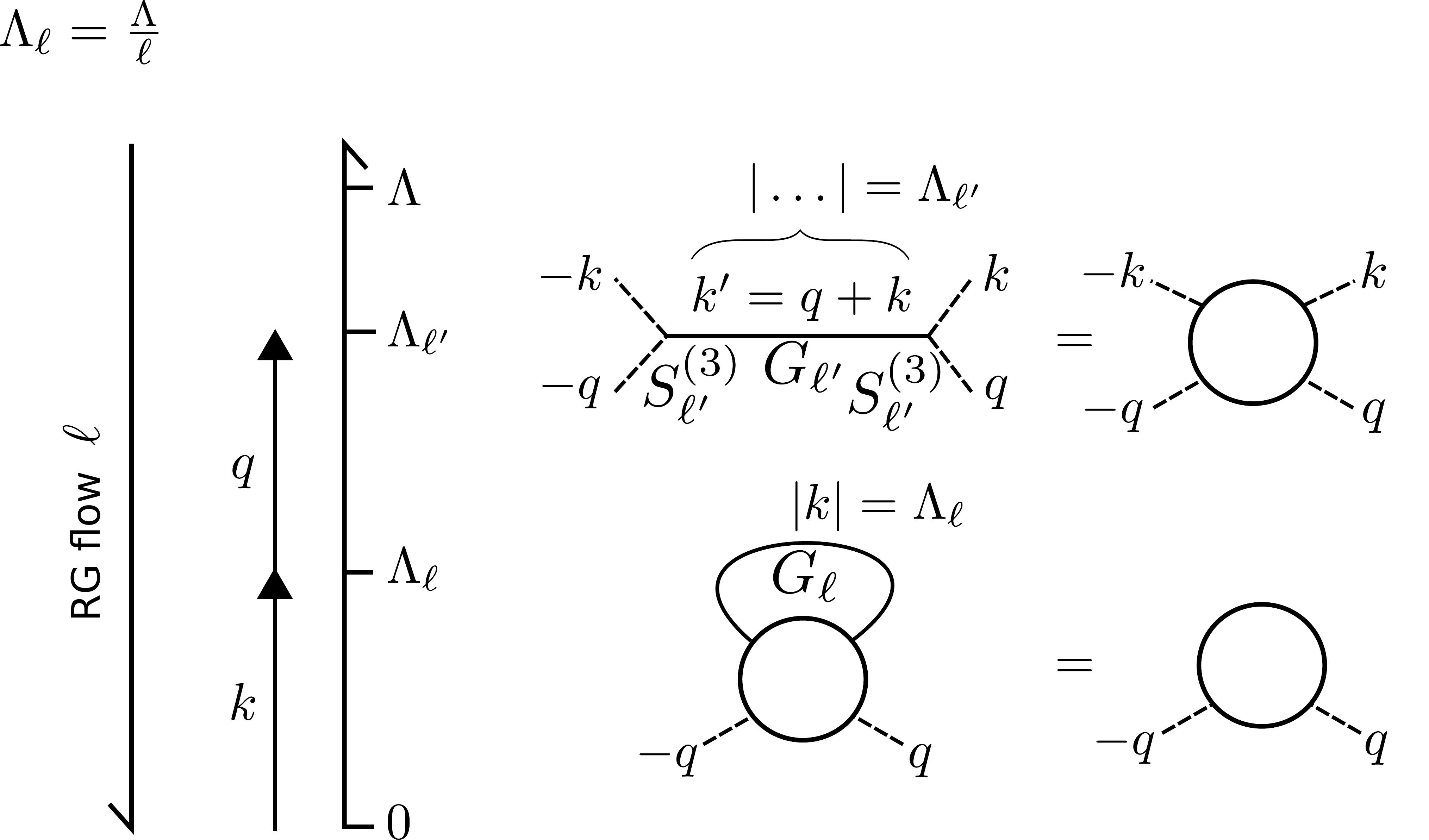}
\par\end{centering}
\caption{\textbf{\textcolor{black}{Flow of momentum dependent vertex caused
by decimation step.}}\textcolor{black}{{} An effective four-point vertex
of reducible tree shape (top) is produced within decimation step $\ell^{\prime}$.
The shell $\ell^{\prime}$ in is determined by the momentum $|k^{\prime}|=|q+k|=\Lambda_{\ell^{\prime}}$
on the propagator line $G_{\ell^{\prime}}$. The propagator $G_{\ell^{\prime}}$
and the three-point vertices $S_{\ell^{\prime}}^{(3)}$ take the values
of the corresponding RG step $\ell^{\prime}$. Later in the flow,
at $\ell>\ell^{\prime}$, two legs of the four-point vertex are contracted
to form a contribution to the two-point vertex $S_{\ell}^{(2)}$ (self-energy).
At fixed and given external momentum $q$, the decimation of shell
$\ell$ thus involves two four-point vertices $S_{\ell^{\prime}}$
and one propagator $G_{\ell^{\prime}}$ that take their values they
had in the earlier shell $\ell^{\prime}<\ell$.\label{fig:Decimation-causes-flow-momentum-1}}}
\end{figure}

In conclusion, the dependence of vertices on external momenta is crucial
to understand the integration of the flow equation. Formally this
means that treating momentum-dependence of vertices requires the solution
of a delay differential equation in $\ell$, where the delay is determined
by the value of the external momenta. Practically, the resulting procedure
is simple:
\begin{itemize}
\item contributions from tree diagrams to the coarse-grained action can
be integrated trivially; they have a non-vanishing value whenever
the momenta of all propagators that connect the vertices of the tree
are not below the current cutoff $\Lambda_{\ell}$;
\item only one-loop diagrams where a propagator connects to a single vertex
contribute non-trivially to the flow; in such diagrams we need to
replace the vertex by all reducible subgraphs, determined by the previous
point; consequently, at non-zero external momenta there is only precisely
one single-scale propagator, a propagator whose momentum is constrained
to the shell of the current decimation step;
\item the structure and combinatorial factors of the diagrams are identical
to the usual one-loop fluctuation expansion; the appearance of precisely
one single-scale propagator in the loop is structurally the same as
in the Wetterich equation.
\end{itemize}
The coupling of different momentum shells arises only if one considers
the dependence of vertices on external, non-zero momenta. As long
as the flow of vertices is computed at vanishing external momenta,
all propagator lines in a one-loop diagram necessarily lie on the
same shell. This special case is treated in most text books and in
the original review by Wilson \cite{Wilson75_773}; it corresponds
to diagrams appearing in the second line of \prettyref{eq:self-energy-WH}.

The combination of vertices that belong to different momentum shells,
at first sight, seems to be in contradiction to the required locality
property of a flow equation, which is of general form $\ell\frac{dK_{\ell}}{d\ell}=\frac{\partial T}{\partial\ell}(K_{\ell},\ell)\big|_{\ell=1}$,
where $K_{\ell}$ is a coupling and $T$ a general coarse-graining
operator \cite{Delamotte12}: the right hand side should only depend
on the couplings $K_{\ell}$ at the current scale $\ell$, not on
the couplings of shells in the RG past $\ell^{\prime}<\ell$; this
property ensures that the semi-group composition law holds. This contradiction,
however, is only apparent because the value of a reducible diagram
does not change except for in the one particular shell $\ell^{\prime}$
which is determined by the momentum $k$ of the connecting propagator,
$|k|=\Lambda_{\ell^{\prime}}$; so the diagram has the same value
at all later coarse-graining steps $\ell>\ell^{\prime}$. If we kept
track of the tree diagrams separately, they would ``automatically''
take the right value if combined into a loop diagram. This would mean,
however, that we follow the flow of the action as a functional\textbf{
}of $\varphi(k)$ for all values of $k$. The direct insertion of
tree diagrams can thus be considered a short cut to alleviate the
book keeping.

\subsection{Decimating an infinitesimal momentum shell\label{sec:Decimating-an-infinitesimal-shell}}

A decimation step of the form \prettyref{eq:self-energy-WH-trees-inserted-1},
where all tree-diagrams have been inserted, requires the computation
of a one-loop integral. We here derive its concrete form. A one-loop
integral across the momentum shell \prettyref{eq:def_shell-1} is
of the form
\begin{eqnarray}
I_{\delta} & := & \int_{\Lambda_{\ell(1+\delta)}<|k|<\Lambda_{\ell}}\,f(k)\label{eq:I_ell}\\
 & = & \frac{1}{(2\pi)^{d}}\,\int_{\Lambda_{\ell(1+\delta)}}^{\Lambda_{\ell}}dk\,k^{d-1}\,\int d\Omega\,f(\Omega\cdot k),\nonumber 
\end{eqnarray}
where $d\Omega$ is the angular integral. Now taking the limit $\delta\to0$
\prettyref{eq:limit_delta_0} we may expand the integral \prettyref{eq:I_ell}
in $\delta$ to obtain
\begin{eqnarray}
 &  & \lim_{\delta\to0}\,\frac{I_{\delta}}{\delta}\label{eq:momentum_shell_expansion}\\
 & = & \lim_{\delta\to0}\,\frac{1}{\delta}\Big(\,\delta\,\ell\,\underbrace{\frac{d\Lambda_{\ell}}{d\ell}}_{-\frac{\Lambda}{\ell^{2}}}\frac{d}{d\Lambda_{a}}\,\frac{1}{(2\pi)^{d}}\int_{\Lambda_{a}}^{\Lambda_{\ell}}dk\,k{}^{d-1}\,\int d\Omega\,f(\Omega\cdot k)\Big|_{\Lambda_{a}=\Lambda_{\ell}}+\o(\delta^{2})\,\Big)\nonumber \\
 & = & \frac{\Lambda_{\ell}^{d}}{(2\pi)^{d}}\,\int d\Omega\,f(\Omega\cdot\Lambda_{\ell}).\nonumber 
\end{eqnarray}
In the following we write for short
\begin{eqnarray*}
\int d\Omega\,f(\Omega\cdot\Lambda_{\ell}) & \equiv & \int_{|k|=\Lambda_{\ell}}f(k).
\end{eqnarray*}
\textbf{}

\subsection{Rescaling of momenta and wavefunction renormalization\label{sec:Rescaling-of-momenta}}

To find fixed points of the RG equations the length scale and hence
the momentum scale are rescaled so that the ranges of momenta are
the same before and after a coarse-graining step; otherwise there
cannot possibly be any fixed points. It is common to define the new
momenta $k_{\ell}$ as \prettyref{eq:def_rescaled_k}. As a consequence
these now span again the same space as did the $k$ prior to any decimation
\begin{eqnarray*}
0< & |k_{\ell}| & <\Lambda.
\end{eqnarray*}
To express the coarse-grained action in the corresponding new field
variables, one defines

\begin{eqnarray}
\varphi_{\ell}(k_{\ell}) & := & \ell^{-1-\frac{d-\eta}{2}}\,\varphi_{<}(q),\label{eq:def_phi_ell}
\end{eqnarray}
where the relation between $k_{\ell}$ and $q$ is fixed by \prettyref{eq:def_rescaled_k}
and we introduced the wavefunction renormalization factor $\ell^{-1-\frac{d-\eta}{2}}$
and the anomalous dimension $\eta$ as usual to keep the $k^{2}$-dependent
coefficient $r^{(2)}$ in \prettyref{eq:phi4_theory} invariant under
rescaling \cite{Wilson75_773}: the factor $\ell^{-1-\frac{d}{2}}$
is chosen such that the Gaussian theory alone would maintain scale-invariance,
as shown in the following:

Our aim is to bring the coarse-grained action  into the same form
as the original Gaussian action  at the expense of modified parameters.
For the Gaussian part of the action (see also \prettyref{sec:Gaussian_model}
for details), expressing $\varphi_{<}$ by the rescaled field $\varphi_{\ell}$
\prettyref{eq:def_phi_ell} yields, with the substitution $\int_{q}\equiv\ell^{-d}\,\int_{k_{\ell}}$,
\begin{eqnarray}
S_{\ell}[\varphi_{<}] & = & \ln\,\Z_{>}-\frac{1}{2}\int_{0\le|q|<\Lambda_{\ell}}\,\varphi_{<}(-q)\,(r^{(0)}+r^{(2)}q^{2})\,\varphi_{<}(q)\label{eq:Gaussian_rescaled}\\
 & = & \ln\,\Z_{>}-\frac{1}{2}\,\ell^{2+d-\eta}\,\ell^{-d}\,\int_{0\le|k_{\ell}|<\Lambda}\,\varphi_{\ell}(-k_{\ell})\,(r^{(0)}+r^{(2)}\,\ell^{-2}\,k_{\ell}^{2})\,\varphi_{\ell}(k_{\ell}).\nonumber 
\end{eqnarray}
So the coefficient of the $k_{\ell}^{2}$-dependent term must be chosen
as
\begin{eqnarray}
r_{\ell}^{(2)} & := & r^{(2)}\,\ell^{-\eta},\label{eq:wavefunction_renorm}
\end{eqnarray}
for the action to maintain the same form. Analogously the parameter
$r_{\ell}^{(0)}$, the coefficient of $\varphi^{2}$, is read off
from \prettyref{eq:Gaussian_rescaled} as
\begin{eqnarray}
r_{\ell}^{(0)} & := & r^{(0)}\,\ell^{2-\eta}.\label{eq:RG_Gaussian}
\end{eqnarray}
Performed infinitesimally, the rescaling step thus contributes
\begin{eqnarray}
\ell\frac{dr_{\ell}^{(0)}}{d\ell} & = & (2-\eta)\,r_{\ell}^{(0)}+\ldots,\label{eq:rescaling_r0}\\
\ell\frac{dr_{\ell}^{(2)}}{d\ell} & = & -\eta\,r_{\ell}^{(2)}+\ldots,\label{eq:rescaling_r2}
\end{eqnarray}
Here the ellipses are the terms from the decimation step.

Correspondingly, the interaction terms $S_{\ell}^{(n>2)}$ transform
under the rescaling

\begin{eqnarray*}
 &  & \int_{q_{1}}\cdots\int_{q_{n-1}}\,S^{(n)}(\{q\})\,\varphi_{<}^{n}(q)\\
 & = & \ell^{n\,\big(1+\frac{d-\eta}{2}\big)}\int_{k_{\ell,1}}\ell^{-d}\cdots\int_{k_{\ell,n-1}}\ell^{-d}\,S^{(n)}(\ell^{-1}\{k_{\ell}\})\,\varphi_{\ell}^{n}(k_{\ell}),
\end{eqnarray*}
so that the coarse-grained coefficients are defined as
\begin{eqnarray}
S_{\ell}^{(n)}(\{k_{\ell}\}) & := & \ell^{n\,\big(1+\frac{d-\eta}{2}\big)-(n-1)d}\,S^{(n)}(\ell^{-1}\{k_{\ell}\})\label{eq:general_vertex_scaling}\\
 & = & \ell^{n\,\big(1-\frac{d+\eta}{2}\big)+d}\,S^{(n)}(\ell^{-1}\{k_{\ell}\}).\nonumber 
\end{eqnarray}
We would like to express the rescaling now for an infinitesimal change
of scale, as we did for \prettyref{eq:rescaling_r0}. We obtain from
\prettyref{eq:general_vertex_scaling}
\begin{eqnarray}
\ell\,\frac{dS_{\ell}^{(n)}(\{k_{\ell}\})}{d\ell} & = & \big(n\,\big(1-\frac{d+\eta}{2}\big)+d\big)\,S_{\ell}^{(n)}(\{k_{\ell}\})\label{eq:rescaling_mom_dep}\\
 & - & k_{\ell}\,\frac{dS_{\ell}^{(n)}(\{k_{\ell}\})}{dk_{\ell}},\nonumber 
\end{eqnarray}
where we used the chain rule to obtain the second term.

Since we perform the coarse-graining and the rescaling for an infinitesimal
step $\delta\to0$, the right hand sides of \prettyref{eq:exact_flow_eq}
and \prettyref{eq:rescaling_mom_dep} simply add up to produce the
final flow equation; cross terms are of order $\o(\delta^{2})$ and
hence drop out in the infinitesimal formulation.

\subsection{Treating momentum dependence in the $\varphi^{4}$-theory\label{sec:Momentum-dependence}}

At the end of \prettyref{sec:Two-classes-of-graphs} the dependence
on external momenta was identified as the reason why different decimation
steps need to be combined to form the one-loop diagrams that cause
the flow in a given shell. In particular, it was shown that the momentum-dependence
leads to a functional delay-differential equation in $\ell$, where
the delay is a function of the external momentum. This section explicitly
treats this momentum dependence on the example of the $\varphi^{4}$
theory \prettyref{eq:phi4_theory} to obtain an improved solution
for the anomalous dimension beyond its leading order approximation
$\eta=0$.

For conceptual clarity, consider the continuous flow equation as a
result of a sequence of decimation and rescaling steps with an infinitesimal
thickness of $\delta$ for each shell, where in the $i$-th step the
flow parameter assumes the value
\begin{eqnarray*}
\ell_{i} & = & 1+i\cdot\delta.
\end{eqnarray*}
We write for short for a momentum $k\in i$ to denote that it is within
the $i$-th shell, namely
\begin{eqnarray*}
\Lambda_{\ell_{i+1}}<|k|<\Lambda_{\ell_{i}}\quad\leftrightarrow & \quad & k\in i.
\end{eqnarray*}

\begin{figure}
\begin{centering}
\includegraphics[width=0.9\textwidth]{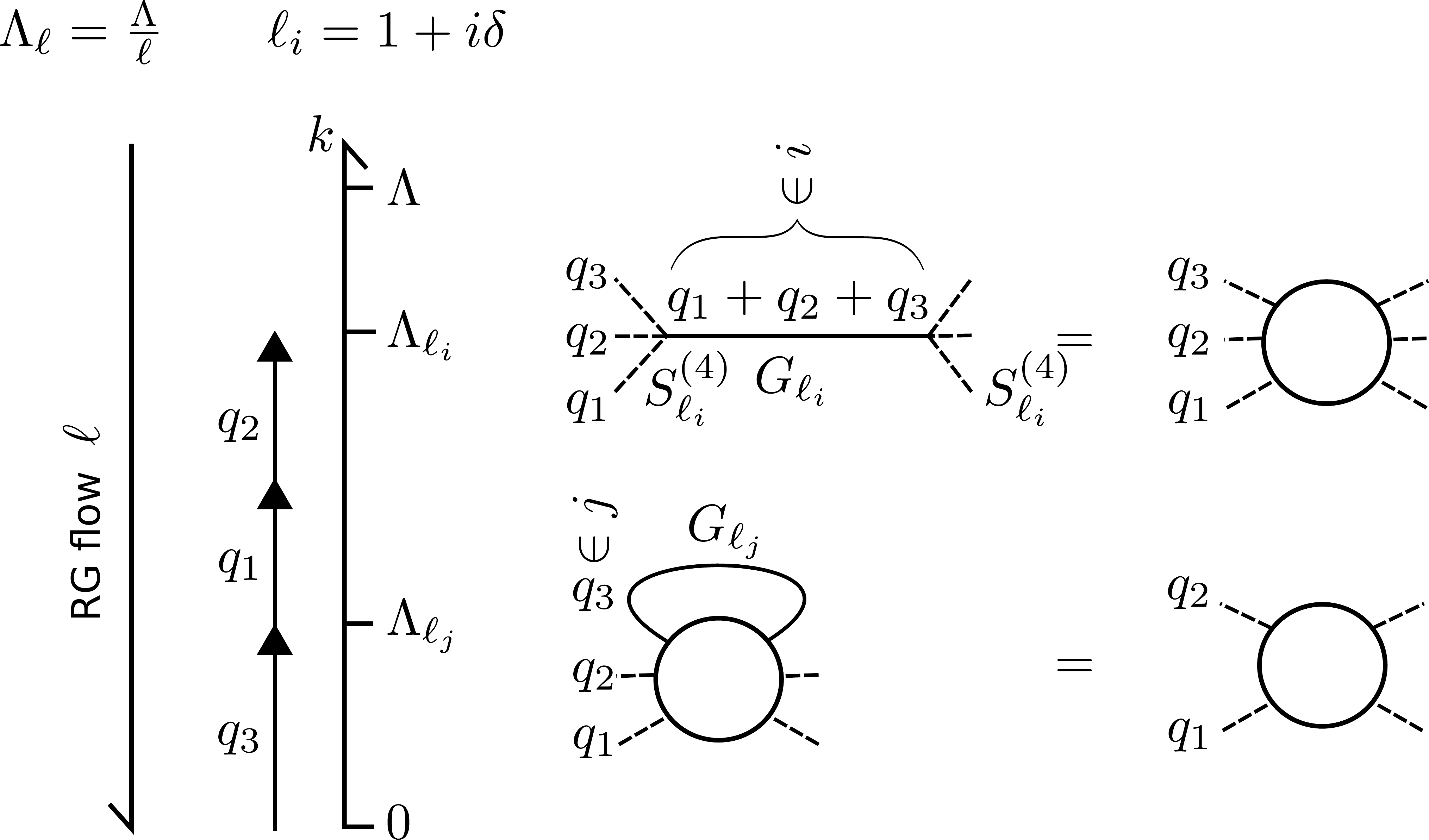}
\par\end{centering}
\caption{\textbf{Flow of momentum dependent vertex from the decimation step.
}An effective six-point vertex of reducible tree shape (top) is produced
within decimation step $i$. The shell $i$ in which the six-point
vertex is produced is determined by the momentum $q_{1}+q_{2}+q_{3}\in i$
on the propagator line $G_{\ell_{i}}$. The propagator $G_{\ell_{i}}$
and the four-point vertices $S_{\ell_{i}}^{(4)}$ take the values
of the corresponding RG step $i$.\textbf{ }Later in the flow, at
$j>i$, two legs of the six-point vertex are contracted to form a
contribution to the four-point vertex $S_{\ell_{j}}^{(4)}$. For given
external momenta $q_{1},q_{2}$, the integration over the loop momentum
$q_{3}$ thus involves six point vertices produced in different shells
$i$.\label{fig:Decimation-causes-flow-momentum}}
\end{figure}

\paragraph{Decimation}

We here investigate the flow for momentum-dependent vertices on the
example of the four point vertex. This is illustrated in \prettyref{fig:Decimation-causes-flow-momentum},
where the flow parameter $\ell=1,\ldots,\infty$ runs from top to
bottom. Each decimation step performs the marginalization of the modes
$\varphi(k)$ that belong to the shell $k\in i$.

Above, we have distinguished two classes of graphs. The tree-shaped
graphs are simple: They contain a single chain of propagators. The
momentum $k$ of the contracted pair of fields belongs to precisely
one decimation step, the step for which $k\in i$. In \prettyref{fig:Decimation-causes-flow-momentum},
the $i$-th step contributes a tree shaped diagram of two four point
vertices; it thus yields a contribution to the six-point vertex, because
there are six uncontracted fields $\varphi_{<}$ left. The value of
the propagator is the one that belongs to the corresponding shell,
because it results from the marginalization of the corresponding modes.
For spherically symmetric propagators, its value is $G_{\ell_{i}}(\Lambda_{\ell_{i}}).$
The entire contribution to the coarse-grained action is therefore
\begin{eqnarray}
S^{(6)}(q_{1},\ldots,q_{6}) & = & 4\cdot4\cdot\frac{1}{2!}\,\left(\frac{S_{\ell_{i}}^{(4)}}{4!}\right)^{2}G_{\ell_{i}}(q_{1}+q_{2}+q_{3}).\label{eq:S6}
\end{eqnarray}
Correspondingly, it holds that
\begin{eqnarray}
q_{1}+q_{2}+q_{3} & \in & i.\label{eq:q_in_i}
\end{eqnarray}
The combinatorial factor $4\cdot4$ caters for the number of possible
legs to choose from to be contracted; the factor $1/2!$ stems from
the appearance of two vertices.

Now consider a contribution to the four-point vertex at non-vanishing
external momenta $q_{1}+q_{2}\neq0$, as shown in \prettyref{fig:Decimation-causes-flow-momentum},
bottom. The contribution is composed of the six-point vertex \prettyref{eq:S6}
with two legs contracted. The decimation step must necessarily be
$j\ge i$, for otherwise the six-point vertex would not exist, so
\begin{eqnarray}
q_{3} & \in & j.\label{eq:q_in_j}
\end{eqnarray}
The contribution to the flow of the four point vertex therefore reads

\begin{eqnarray}
\label{eq:correction_four_point}\\
\ell\frac{1}{4!}\frac{dS_{\ell}^{(4)}}{d\ell} & = & \underbrace{4\cdot4\cdot3\cdot3\cdot\frac{1}{2}\cdot\frac{1}{2!}}_{=36}\,\frac{1}{(2\pi)^{d}}\,\int_{|q_{3}|=\Lambda_{j}}\left(\frac{S_{\ell_{i}}^{(4)}}{4!}\right)^{2}G_{\ell_{i}}(q_{1}+q_{2}+q_{3})\,G_{\ell_{j}}(q_{3}),\nonumber 
\end{eqnarray}
where the combinatorial factor $3\cdot3\cdot\frac{1}{2}$ represents
the combinations of selecting a pair of the legs of the six point
vertex from the two sets of three legs each. The factor $1/2$ has
the same reason as in an $n$ choose $2$ factor: the order by which
we pick the pair does not matter, so we need to correct for this factor,
since we otherwise double-count contractions. The combinatorial factor
in total is $4\cdot4\cdot3\cdot3\cdot\frac{1}{2}=72$, identical to
that of the usual one loop diagram.

\paragraph{Rescaling}

\begin{figure}
\begin{centering}
\includegraphics[width=0.5\textwidth]{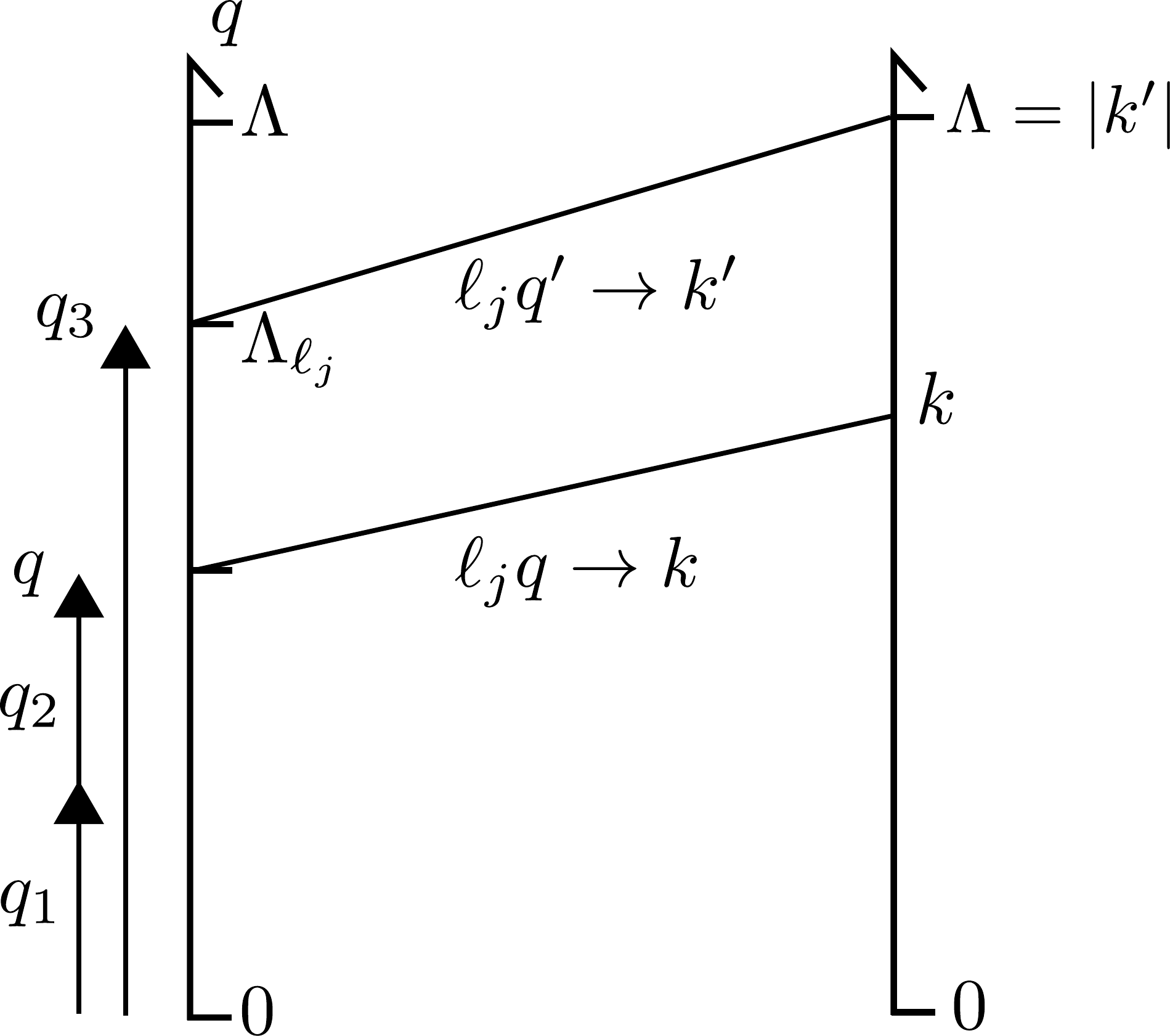}
\par\end{centering}
\caption{\textbf{Rescaling of momenta. }Momenta \textbf{$q$ }before rescaling.
Here $q_{1}$ and $q_{2}$ are external momenta and $q_{3}=q^{\prime}$
is the loop momentum in \prettyref{eq:correction_four_point}. Momenta
$k=\ell_{j}q$ after rescaling. The rescaled loop momentum $k^{\prime}=\ell_{j}q^{\prime}$
is located at the cutoff of the decimation step, $|k^{\prime}|=\Lambda$.\label{fig:Rescaling-of-momenta}}
\end{figure}

To obtain the scaling form of the flow equation, we need to perform
the rescaling of fields and momenta, as described in \prettyref{sec:Rescaling-of-momenta}.

The six point vertex is generated in the step $i<j$. In the rescaled
units, the momenta in step $i$ and step $j$ obey, with \prettyref{eq:def_rescaled_k},
the relation
\begin{eqnarray*}
k & = & \ell_{j}\,q.
\end{eqnarray*}
The six point vertex, expressed in terms of the quantities at the
scale $\ell_{j}$, with \prettyref{eq:general_vertex_scaling}, is

\begin{eqnarray*}
S_{\ell_{j}}^{(n)}(\{k\}) & = & \ell_{j}^{n\,\big(1-\frac{d+\eta}{2}\big)+d}\,S^{(n)}(\ell_{j}^{-1}\{k\}).
\end{eqnarray*}
Conversely, we need to express $S^{(n)}$ in terms of the rescaled
coefficient at scale $i$, for which the same relation holds, but
with $i\leftrightarrow j$, so that we get in total
\begin{eqnarray*}
S_{\ell_{j}}^{(n)}(\{k\}) & = & \left(\frac{\ell_{j}}{\ell_{i}}\right)^{n\,\big(1-\frac{d+\eta}{2}\big)+d}\,S_{\ell_{i}}^{(n)}(\frac{\ell_{i}}{\ell_{j}}\,\{k\}).
\end{eqnarray*}
So only the ratio $\ell_{i}/\ell_{j}$ of the coarse-graining scale
parameter appears. This ratio, in turn, is related to the momenta
by \prettyref{eq:q_in_i} and \prettyref{eq:q_in_j} with $q:=q_{1}+q_{2}$
and $q^{\prime}:=q_{3}$ as $|q+q^{\prime}|=\frac{\Lambda}{\ell_{i}}$
and $|q^{\prime}|=\frac{\Lambda}{\ell_{j}}$ we have
\begin{eqnarray}
\frac{\ell_{j}}{\ell_{i}} & = & \frac{|q^{\prime}+q|}{|q^{\prime}|}.\label{eq:ratio_momenta_scales}
\end{eqnarray}
Here the momenta $q$ and $q^{\prime}$ are measured at the absolute
scale, as shown in \prettyref{fig:Decimation-causes-flow-momentum}
\textemdash{} they are the arguments of $\varphi_{<}(q)$. But we
want to know how the four point vertex depends on the momenta of the
fields $\varphi_{\ell_{j}}(k)$. We thus need to express the momentum
$q$ that appears in the latter factor by the $k$ at scale $\ell_{j}$,
as illustrated in \prettyref{fig:Rescaling-of-momenta}. The coarse-graining
scale $\ell_{j}$ satisfies $\ell_{j}|q^{\prime}|\equiv\ell_{j}|q_{3}|=\Lambda$.
At this scale, the momenta $q$ take the value $\ell_{j}q=:k$, so
\begin{eqnarray*}
q & = & \frac{1}{\ell_{j}}\,k=\frac{|q^{\prime}|}{\Lambda}\,k,\\
q^{\prime} & = & \frac{1}{\ell_{j}}\,k^{\prime}=\frac{|q^{\prime}|}{\Lambda}\,k^{\prime}.
\end{eqnarray*}
We thus have
\begin{eqnarray}
\left(\frac{\ell_{j}}{\ell_{i}}\right)^{2} & = & \frac{(q^{\prime}+q)^{2}}{q^{\prime2}}=\frac{(k^{\prime}+k)^{2}}{\Lambda^{2}}.\label{eq:l_i_div_l_j}
\end{eqnarray}
Expressed in rescaled variables, the correction \prettyref{eq:correction_four_point}
to the four-point vertex hence takes the form

\begin{eqnarray}
 &  & 3\cdot3\cdot\frac{1}{2}\cdot\,\frac{1}{(2\pi)^{d}}\int_{|k^{\prime}|=\Lambda}\,\theta(\ell_{i}-\ell_{j}+)\,\left(\frac{\ell_{j}}{\ell_{i}}\right)^{6\,\big(1-\frac{d+\eta}{2}\big)+d}\,\frac{S_{\ell_{i}}^{(6)}\big(\frac{\ell_{i}}{\ell_{j}}\,\{k\}\big)}{6!}\,G_{\ell_{j}}(k^{\prime})\label{eq:correction_four_point-2}\\
 & = & \underbrace{36}_{4\cdot4\cdot3\cdot3\cdot\frac{1}{2}\cdot\frac{1}{2!}}\cdot\frac{1}{(2\pi)^{d}}\,\int_{|k^{\prime}|=\Lambda}\,\theta(\ell_{i}-\ell_{j}+)\,\left(\frac{\ell_{j}}{\ell_{i}}\right)^{6\,\big(1-\frac{d+\eta}{2}\big)+d}\,\left(\frac{S_{\ell_{i}}^{(4)}\big(\frac{\ell_{i}}{\ell_{j}}\,\{k\}\big)}{4!}\right)^{2}G_{\ell_{i}}(\Lambda)\,G_{\ell_{j}}(\Lambda)\nonumber \\
 & \simeq & 36\cdot\frac{1}{(2\pi)^{d}}\,\left(\frac{S_{\ast}^{(4)}}{4!}\right)^{2}\,\int_{|k^{\prime}|=\Lambda}\,1+\theta(k^{\prime}\cdot k+)\,\Big(6-2d-3\eta\Big)\times\label{eq:correction_four_point_full}\\
 &  & \times\Bigg(\,\frac{(k^{\prime}\cdot k)}{\Lambda^{2}}+\frac{k^{2}}{2\,\Lambda^{2}}+\frac{(k^{\prime}\cdot k)^{2}}{\Lambda^{4}}\,\Big(2-d-\frac{3}{2}\eta\Big)\Bigg)\,G_{\ell_{i}}(\Lambda)\,G_{\ell_{j}}(\Lambda).\nonumber 
\end{eqnarray}
From the second to the third line we expanded the term $\left(\frac{\ell_{j}}{\ell_{i}}\right)^{6\,\big(1-\frac{d+\eta}{2}\big)+d}$
up to second order in $k^{\prime}$ by using \prettyref{eq:l_i_div_l_j}
(see \prettyref{sec:Expansion-of-momentum-dependence} for details).
The notation $\theta(x+)$ is to be read as $\lim_{\epsilon\searrow0}\theta(x+\epsilon)$.
The Heaviside functions ensure that the effective six-point vertex
has already been produced at the decimation step $i$, namely that
$\ell_{i}\ge\ell_{j}$. In the last line, we rewrite this condition,
amounting to $|k+k^{\prime}|>|k^{\prime}|$ as $k$ and $k^{\prime}$
having a parallel component, $\theta(k^{\prime}\cdot k+)$.

The four point vertices $S_{\ell_{i}}^{(4)}$ and the propagator $G_{\ell_{i}}$
that appear here are those expressed in terms of the rescaled variables
at scale $\ell_{i}$, because they take the values they had in the
$i$-th decimation step. The momentum of each propagator in rescaled
units is always equal to the original cutoff. In the last step we
assumed that the flow will be close to the fixed point. As a result,
we may replace the vertices and the propagators by their values at
the fixed point, denoted by a star. 

In addition, in the last step we neglected the effect of the momentum
dependence of $S^{(4)}$ on its own corrections. We will a posteriori
see that the neglected terms are at most of order $\big(u_{\ast}^{(0)}\big)^{3}$,
where $u_{\ast}^{(0)}$ is the momentum-independent part of the interaction.
This approximation is good as long as the interaction is small. In
principle, the momentum-dependence of $S^{(4)}$ can be taken into
account at the expense of additional computations.

If we assumed $d=4$ and $\eta=0$, the last line in \prettyref{eq:correction_four_point_full}
would simplify to

\begin{eqnarray}
 &  & 36\cdot\frac{1}{(2\pi)^{4}}\,\int_{|k^{\prime}|=\Lambda}\frac{\theta(k^{\prime}\cdot k+)\,\Lambda^{2}}{(k^{\prime}+k)^{2}}\,\left(\frac{S_{\ast}^{(4)}}{4!}\right)^{2}G_{\ast}(\Lambda)\,G_{\ast}(\Lambda).\label{eq:approximate_four_point_flow}
\end{eqnarray}
This approximation is justified if one wants to compute the correction
only in $d=4-\epsilon$ dimensions, in which case $\eta\propto\epsilon^{2}$,
as will be seen in the following. In the following we do not make
this approximation, but rather keep the full dependence on $d$ and
$\eta$ in \prettyref{eq:correction_four_point_full}.

The main insight of this section is the appearance of the first factor
in the last line of \prettyref{eq:correction_four_point_full} that,
by \prettyref{eq:q_in_i}, arises from the rescaling term in combination
with the momentum-dependence of the shell index $\ell_{i}(q_{1}+q_{2}+q_{3})$.
The latter implies a dependence on the external momenta $q_{1}+q_{2}$
of the effective four-point vertex. It is a direct consequence of
the insertion of the tree diagram produced in the decimation step
$\ell_{i}$ previous to the current one, $\ell_{j}$; showing that
we solve a delayed functional differential equation here. This momentum-dependence
has been neglected in the approximation made in \cite{Wegner73_401},
as discussed below.

\subsection{Flow of the momentum dependent interaction vertex\label{sec:Flow-of-the-momentum-dependent-interaction}}

We are interested in the long-distance behavior for which we want
to get a flow equation. This requires the momentum dependence of \prettyref{eq:correction_four_point-2}
for small $k$.

We here study the case that the dimensionality of the system is not
close to the upper critical dimension, thus $d\neq4-\epsilon$. One
then cannot make the approximation in \prettyref{eq:approximate_four_point_flow},
but rather needs to use the full expression \prettyref{eq:correction_four_point_full}.
The computation is, however, not more complicated than the approximation,
because for up to order $\mathcal{O}(k^{2})$ the same integrals appear.
These are (details are given in the Appendix \prettyref{sec:Momentum-dependent-integrals-of-four-point},
\prettyref{eq:uniaxial_1}, \prettyref{eq:uniaxial_2})
\begin{eqnarray*}
\int_{|k^{\prime}|=\Lambda}\,\theta(k^{\prime}\cdot k+)\,(k^{\prime}\cdot k) & = & S_{d-1}\,\Lambda\,||k||\,\frac{1}{d-1},\\
\int_{|k^{\prime}|=\Lambda}\,\theta(k^{\prime}\cdot k+)\,k^{2} & = & \frac{S_{d}}{2}\,k^{2},\\
\int_{|k^{\prime}|=\Lambda}\,\theta(k^{\prime}\cdot k+)\,(k^{\prime}\cdot k)^{2} & = & \frac{S_{d}}{2d}\,\Lambda^{2}\,k^{2}.
\end{eqnarray*}
We thus obtain from the decimation step and \prettyref{eq:correction_four_point_full}

\begin{eqnarray}
 &  & \ell\frac{d}{d\ell}\,\frac{1}{4!}S_{\ell}^{(4)}(k_{1},k_{2},k_{3},k_{4})\label{eq:decimation_S4_non_approx}\\
 & = & 36\,\frac{\Lambda^{d}}{(2\pi)^{d}}\,(u_{\ell}^{(0)})^{2}G_{\ell}^{2}(\Lambda)\,\Bigg[S_{d}+\Big(6-2d-3\eta\Big)\times\nonumber \\
 &  & \times\Bigg(\frac{S_{d-1}}{d-1}\,\frac{||k_{1}+k_{2}||}{\Lambda}+\frac{S_{d}}{4d}\,\Big(4-d-3\eta\Big)\,\frac{(k_{1}+k_{2})^{2}}{\Lambda^{2}}\Bigg)\Bigg].\nonumber 
\end{eqnarray}
Had we taken the momentum-dependence of $S^{(4)}$ into account
(see comment after \prettyref{eq:correction_four_point_full}), we
would have gotten additional terms here. We will a posteriori see
that the neglected terms are of order $\o(\big(u_{\ast}^{(0)}\big)^{3})$
and discuss their importance close to $d=3$ after eq. \prettyref{eq:rel_u1}.

So the renormalized action requires a momentum-dependent $S^{(4)}$-interaction
of the form
\begin{eqnarray}
-\frac{1}{4!}S_{\ell}^{(4)}(k_{1},k_{2},k_{3},k_{4}) & = & u_{\ell}^{(0)}+u_{\ell}^{(1)}\,||k_{1}+k_{2}||+u_{\ell}^{(2)}\,(k_{1}+k_{2})^{2}\label{eq:S4_momentum_dep}\\
 & =: & \Diagram{gd & p & gu\\
gu &  & gd
}
.\nonumber 
\end{eqnarray}
Defining the rescaled parameters with subscript $\ell$, according
to \prettyref{eq:general_vertex_scaling} with $n=4$ yields

\begin{eqnarray}
-\frac{1}{4!}S_{\ell}^{(4)}(k_{\ell1},k_{\ell2}) & =: & u_{\ell}^{(0)}+u_{\ell}^{(1)}\,||k_{\ell1}+k_{\ell2}||+u_{\ell}^{(2)}\,\big(k_{\ell1}+k_{\ell2}\big)^{2}\label{eq:momentun_dependent_S4-1}\\
 & \stackrel{(\ref{eq:general_vertex_scaling})}{=} & \ell^{4-d-2\eta}\,\big(u^{(0)}+u^{(1)}\,\ell^{-1}||k_{\ell1}+k_{\ell2}||+u^{(2)}\,\ell^{-2}\,\big(k_{\ell1}+k_{\ell2}\big)^{2}\big).\nonumber 
\end{eqnarray}
In differential form the contribution from rescaling is thus
\begin{eqnarray}
\ell\frac{d}{d\ell}u_{\ell}^{(0)} & = & (4-d-2\eta)\,u_{\ell}^{(0)}+\ldots,\label{eq:rescaling_u0}\\
\ell\frac{d}{d\ell}u_{\ell}^{(1)} & = & (3-d-2\eta)\,u_{\ell}^{(1)}+\ldots,\label{eq:rescaling_u1}\\
\ell\frac{d}{d\ell}u_{\ell}^{(2)} & = & (2-d-2\eta)\,u_{\ell}^{(2)}+\ldots,\label{eq:rescaling_u2}
\end{eqnarray}
where the momentum-dependent term $k_{\ell}\,\frac{dS_{\ell}^{(n)}(k_{\ell})}{dk_{\ell}}$
in \prettyref{eq:rescaling_mom_dep} is taken care of by explicitly
rescaling the momenta appearing in \prettyref{eq:momentun_dependent_S4-1}.

The complete flow equations of the parameters $u_{\ell}^{(0)}$, $u_{\ell}^{(1)}$,
and $u_{\ell}^{(2)}$ then follow by including the decimation step
\prettyref{eq:decimation_S4_non_approx} as
\begin{eqnarray}
\ell\frac{du_{\ell}^{(0)}}{d\ell} & = & (4-d-2\eta)\,u_{\ell}^{(0)}\label{eq:flow_eq_u0}\\
 &  & -36\,S_{d}\,\frac{\Lambda^{d}}{(2\pi)^{d}}\,\frac{(u_{\ell}^{(0)})^{2}}{(r_{\ell}^{(0)}+r_{\ell}^{(2)}\Lambda^{2})^{2}},\nonumber \\
\nonumber \\
\ell\frac{du_{\ell}^{(1)}}{d\ell} & = & (3-d-2\eta)\,u_{\ell}^{(1)}\label{eq:flow_eq_u1}\\
 &  & -36\,\big(6-2d-3\eta\big)\,\frac{S_{d-1}}{d-1}\,\frac{\Lambda^{d}}{(2\pi)^{d}}\,\frac{(u_{\ell}^{(0)})^{2}}{(r_{\ell}^{(0)}+r_{\ell}^{(2)}\Lambda^{2})^{2}}\,\frac{1}{\Lambda},\nonumber \\
\nonumber \\
\ell\frac{du_{\ell}^{(2)}}{d\ell} & = & (2-d-2\eta)\,u_{\ell}^{(2)}\label{eq:flow_eq_u2}\\
 &  & -36\,\big(6-2d-3\eta\big)\,\big(4-d-3\eta\big)\,\frac{S_{d}\,\Lambda^{d}}{4d\,(2\pi)^{d}}\,\frac{(u_{\ell}^{(0)})^{2}}{(r_{\ell}^{(0)}+r_{\ell}^{(2)}\Lambda^{2})^{2}}\,\frac{1}{\Lambda^{2}}.\nonumber 
\end{eqnarray}
The flow equation for $u_{\ell}^{(2)}$ has the property that for
$d=3$ the first factor in the second row is $-3\eta$ and for $d=4$,
the second factor becomes $-3\eta$; this means that the quadratic
momentum dependence in both cases becomes small, $u_{\ast}^{(2)}/(u_{\ast}^{(0)})^{2}\propto\eta\,$.
In between these two limits, the corrections are larger. The flow
equation for $u_{\ell}^{(1)}$ has a first factor that is $\propto-3\eta$
for $d=3$. But at the same time, the rescaling term becomes $-2\eta$,
so $\eta$ drops out of the fixed point equation. The $|k|$-dependence
of the interaction will thus dominate over the $k^{2}$-dependence
in $d=3$ dimensions. This is an example that shows that the non-differentiable
$k$-dependence may become important.

By the rescaling, we express propagators and vertices in the correction
terms by the ones at the initial scale, as given by \prettyref{eq:rescaling_mom_dep}.
This also replaces $\Lambda_{\ell}\to\Lambda$, because the rescaling
after each decimation step restores the original range of momenta
and thus keeps the position of the cutoff constant.

\subsection{Flow of the momentum-dependent self-energy\label{sec:Flow-of-the-self-energy}}

Inserting the momentum-dependent term of the interaction $S_{\ell}^{(4)}$
\prettyref{eq:S4_momentum_dep} into the self-energy corrections,
this interaction vertex induces a dependence on $k$. We have two
possibilities of inserting the interaction into the self-energy diagram

\begin{eqnarray*}
 & \Diagram{\vertexlabel^{\varphi_{<}(q)}\\
gd & pf0flfluf0\\
\vertexlabel_{\varphi_{<}(-q)}gu
}
 & =\varphi_{<}(q)\,\varphi_{<}(-q)\,\frac{4\cdot3}{2}\,\frac{\Lambda^{d}}{(2\pi)^{d}}\,\int_{|k|=\Lambda}\,\frac{1}{4!}\,S_{\ell}^{(4)}(-q,q,k,-k)\,G_{\ell}(k),\\
\end{eqnarray*}
which leads to the second and third line of the following equation

\begin{eqnarray}
\ell\frac{d}{d\ell}\,\frac{1}{2}S_{\ell}^{(2)}(q,-q) & := & -2\cdot6\cdot\Diagram{\vertexlabel^{\varphi_{<}(q)}\\
gd & pf0flfluf0\\
\vertexlabel_{\varphi_{<}(-q)}gu
}
\label{eq:correction_self_energy_infinitesimal_anomalous-1}\\
\nonumber \\
\nonumber \\
 & = & -2\cdot2\cdot\frac{\Lambda_{\ell}^{d}}{(2\pi)^{d}}\cdot\int_{|k|=\Lambda_{\ell}}\,\frac{1}{4!}S_{\ell}^{(4)}(q,-q,k,-k)\,G_{\ell}(k)\nonumber \\
 &  & -2\cdot4\cdot\frac{\Lambda_{\ell}^{d}}{(2\pi)^{d}}\cdot\int_{|k|=\Lambda_{\ell}}\,\frac{1}{4!}S_{\ell}^{(4)}(q,k,-k,-q)\,G_{\ell}(k).\nonumber 
\end{eqnarray}
The penultimate line corresponds to the $2$ ways of choosing either
the left pair or the right pair of amputated legs of the interaction
vertex $S^{(4)}$ to be the amputated lines of the self-energy diagram
\textendash{} in both cases we have $q_{1}=q$ and $q_{2}=-q$, so
that the momentum dependence of the four point vertex $S^{(4)}(q,-q,k,-k)$,
according to eq. \prettyref{eq:S4_momentum_dep}, drops out. In the
latter row the four point vertex is inserted such that one external
amputated leg connects to left side of $S^{(4)}$ ($2$ possibilities)
and one to the right side (another $2$ possibilities), so we get
a term 
\begin{eqnarray}
\frac{1}{4!}S_{\ell}^{(4)}(q,k,-k,-q) & = & u_{\ell}^{(0)}+u_{\ell}^{(1)}\,\sqrt{(q+k)^{2}}+u_{\ell}^{(2)}\,(q+k)^{2},\label{eq:q_squared_S4}
\end{eqnarray}
which therefore is dependent on the momentum $q$. Dropping terms
of order $\o(q^{3})$ and higher and sorting the result into terms
$\propto q^{0}$ and $\propto q^{2}$ according to $-S^{(2)}(q)=r_{\ell}^{(0)}+r_{\ell}^{(2)}\,q^{2}+\o(q^{3})$
we get the pair of flow equations (see \prettyref{sec:Momentum-dependence-of-self-energy}
for details)

\begin{eqnarray}
\ell\frac{d}{d\ell}r_{\ell}^{(0)} & = & (2-\eta)\,r_{\ell}^{(0)}\label{eq:flow_eq_r0}\\
 &  & +2\cdot\frac{S_{d}\,\Lambda^{d}}{(2\pi)^{d}}\cdot\frac{6\,u_{\ell}^{(0)}+4\,u_{\ell}^{(1)}\,\Lambda+4\,u_{\ell}^{(2)}\,\Lambda^{2}}{r_{\ell}^{(0)}+r_{\ell}^{(2)}\Lambda^{2}},\nonumber \\
\ell\frac{d}{d\ell}r_{\ell}^{(2)} & = & -\eta\,r_{\ell}^{(2)}\label{eq:flow_eq_r2}\\
 &  & +8\cdot\frac{S_{d}\,\Lambda^{d}}{(2\pi)^{d}}\cdot\frac{\frac{u_{\ell}^{(1)}}{2\Lambda}\,\Big(1-\frac{1}{d}\Big)+u_{\ell}^{(2)}}{r_{\ell}^{(0)}+r_{\ell}^{(2)}\Lambda^{2}},\nonumber 
\end{eqnarray}
where the first line in each flow equation comes from the rescaling
given by \prettyref{eq:rescaling_r0} and \prettyref{eq:rescaling_r2}.
The momentum dependence of the propagator $G_{\ell}(q)\equiv\big(-S^{(2)}(q)\big)^{-1}$
here follows from the momentum dependence of $S^{(2)}(q)$; so only
a constant and a $k^{2}$-dependent term arises in the inverse propagator.

\subsection{Fixed points and critical exponents for $3\le d\le4$\label{sec:Fixed-points-and-critical-exponents}}

We may now compute the Wilson-Fisher fixed point for arbitrary dimensions
between $3$ and $4$. We determine $\eta$ such that $r_{\ast}^{(2)}\equiv1$
is a fixed point of \prettyref{eq:flow_eq_r2}. Throughout we approximate
the mass to be small compared to the cutoff, $r_{\ast}^{(0)}\ll\Lambda^{2}$.
The details can be found in \prettyref{sec:Detailed-calculation-of-FP}.

\subsubsection*{Interaction}

The fixed point for the momentum-independent four point coupling obeys

\begin{eqnarray}
\Lambda^{d-4}\,u_{\ast}^{(0)} & \simeq & (2\pi)^{d}\,\frac{4-d-2\eta}{36\,S_{d}}.\label{eq:u_0_fixed}
\end{eqnarray}
In the limits $d\to3$ and $d\to4$ we get
\begin{eqnarray}
\Lambda^{d-4}\,u_{\ast}^{(0)} & \simeq & \left\{ \begin{array}{cc}
\pi^{2}\frac{1-2\eta}{18}\simeq\frac{\pi^{2}}{18} & d=3\\
\pi^{2}\,\frac{2}{9}\big(\epsilon-2\eta\big) & d=4-\epsilon
\end{array}\right..\label{eq:u_0_3d_4d}
\end{eqnarray}
As explained by Wilson \cite[section V]{Wilson75_773}, all fixed
point values can be related back to the value of the interaction $u_{\ast}^{(0)}$,
the single marginal coupling of the Gell-Mann-Low theory. The first-order
momentum dependence is quadratic in $u_{\ast}^{(0)}$
\begin{eqnarray}
\Lambda^{5-d}\,\frac{u_{\ast}^{(1)}}{\big(u_{\ast}^{(0)}\big)^{2}} & \simeq & \frac{36}{(2\pi)^{d}}\,\frac{2(3-d)-3\eta}{3-d-2\eta}\,\frac{S_{d-1}}{d-1}.\label{eq:rel_u1}
\end{eqnarray}
As long as $u_{0}$ is small, this justifies the approximation made
in the last line of \prettyref{eq:correction_four_point_full}, the
neglect of the momentum-dependence of the four-point vertex. Close
to $d=4$ dimensions this approximation is very good, as expected.
As $d\to3$ dimensions are approached, the error increases, because
with \prettyref{eq:u_0_3d_4d} $\Lambda^{d-4}\,u_{\ast}^{(0)}\simeq\pi^{2}/18\simeq1/2$.

In the limits $d\to3$ and $d\to4$ we get
\begin{eqnarray}
\label{eq:u1_fixed_3d_4d}\\
\Lambda^{d-3}\,u_{\ast}^{(1)} & \simeq & \left\{ \begin{array}{cc}
\frac{\pi^{2}}{48}\,(1-2\eta)^{2}\simeq\frac{\pi^{2}}{48} & d=3\\
\frac{4}{9}\,\frac{2(1+\epsilon)-3\eta}{1+\epsilon-2\eta}\,(\epsilon-2\eta)^{2}\frac{\pi}{(3-\epsilon)}\simeq\frac{8\pi}{27}\,\epsilon^{2}+\mathcal{O}(\epsilon^{3}) & d=4-\epsilon
\end{array}\right..\nonumber 
\end{eqnarray}
The quadratic momentum dependence of the interaction is
\begin{eqnarray}
\Lambda^{6-d}\,\frac{u_{\ast}^{(2)}}{(u_{\ast}^{(0)})^{2}} & \simeq & 36\,\frac{S_{d}}{(2\pi)^{d}}\,\frac{2(3-d)-3\eta}{2-d-2\eta}\,\frac{4-d-3\eta}{4d}.\label{eq:rel_u2}
\end{eqnarray}

As expected from the $\epsilon$-expansion, the fixed-point value
depends quadratically on $u_{\ast}^{(0)}$. For $d\to3$ and $d\to4$
we get

\begin{eqnarray}
\label{eq:u2_fixed_3d_4d}\\
\Lambda^{d-2}u_{\ast}^{(2)} & \simeq & \left\{ \begin{array}{cc}
\frac{\pi^{2}}{72}\,\frac{\eta}{1+2\eta}\,(1-3\eta)(1-2\eta+4\eta^{2})\simeq\frac{(2\pi)^{3}}{72}\,\eta+\mathcal{O}(\eta^{2}) & d=3\\
\frac{2\pi^{2}}{9}\,\frac{2(1-\epsilon)+3\eta}{2-\epsilon+2\eta}\,\frac{(\epsilon-3\eta)(\epsilon-2\eta)^{2}}{4(4-\epsilon)}\simeq\frac{\pi^{2}}{72}\,\epsilon^{3}+\o(\epsilon^{4}) & d=4-\epsilon
\end{array}\right..\nonumber 
\end{eqnarray}

\subsubsection*{Mass term}

To check if the approximation $r_{\ast}^{(0)}\ll\Lambda^{2}$ is justified,
one observes that the fixed point value for $r_{\ast}^{(0)}$, in
this approximation, is determined as

\begin{eqnarray}
\Lambda^{-2}r_{\ast}^{(0)} & \simeq & -\frac{6\,S_{d}}{(2\pi)^{d}}\cdot\,\Lambda^{d-4}u_{\ast}^{(0)}\label{eq:r0_fixed}\\
 & - & \frac{144\,S_{d}}{(2\pi)^{2d}}\,\frac{2(3-d)-3\eta}{3-d-2\eta}\,\Big(\frac{S_{d-1}}{d-1}+\frac{4-d-3\eta}{4d}\,S_{d}\Big)\,(\Lambda^{d-4}u_{\ast}^{(0)})^{2},\nonumber 
\end{eqnarray}
which shows the approximate linear relationship between $r_{\ast}^{(0)}$
and $u_{\ast}^{(0)}$ and explains why the mass term stays small as
long as $u_{\ast}^{(0)}$ is small. In this case, it is justified
to approximate $\Lambda^{2}+r_{\ast}^{(0)}\rightarrow r_{\ast}^{(0)}$.
The ratio between mass and coupling term is

\begin{eqnarray}
\Lambda^{2-d}\frac{r_{\ast}^{(0)}}{u_{\ast}^{(0)}} & \simeq & -\frac{6\,S_{d}}{(2\pi)^{d}}\label{eq:rel_r0}\\
 & - & \frac{4}{(2\pi)^{d}}\,\frac{(2(3-d)-3\eta)(4-d-2\eta)}{3-d-2\eta}\,\Big(\frac{S_{d-1}}{d-1}+\frac{4-d-3\eta}{4d}\,S_{d}\Big).\nonumber 
\end{eqnarray}
The correction to the linear relation in the second line is $\propto\eta$
for $d=4$, but becomes notable in for $d<3$, as shown in \prettyref{fig:Fixed-point-crit-exp}c.
Inserting the expression for $u_{\ast}^{(0)}$ from \prettyref{eq:u_0_fixed}
yields $r_{\ast}^{(0)}$ as a function of $d$, shown in \prettyref{fig:Fixed-point-crit-exp}a.

\paragraph*{Critical exponent $\eta$}

Demanding stationarity of \prettyref{eq:flow_eq_r2} and $r_{\ell}^{(2)}\stackrel{!}{=}1$
we get
\begin{eqnarray}
\eta & = & 8\cdot\frac{S_{d}}{(2\pi)^{d}}\cdot\Big(\frac{1}{2}\Lambda^{d-3}u_{\ast}^{(1)}\,\Big(1-\frac{1}{d}\Big)+\Lambda^{d-2}u_{\ast}^{(2)}\Big).\label{eq:eta_base_eq}
\end{eqnarray}
Expressed in terms of $u_{\ast}^{(0)}$

\begin{eqnarray}
\Lambda^{8-2d}\,\frac{\eta}{(u_{\ast}^{(0)})^{2}} & \simeq & 8\,\frac{S_{d}}{(2\pi)^{2d}}\cdot\Big(\frac{1}{2}\Big(1-\frac{1}{d}\Big)\,36\,\frac{2(3-d)-3\eta}{3-d-2\eta}\,\frac{S_{d-1}}{d-1}\label{eq:rel_eta}\\
 &  & \phantom{8\,\frac{S_{d}}{(2\pi)^{2d}}\cdot}+36\,\frac{2(3-d)-3\eta}{2-d-2\eta}\,\frac{4-d-3\eta}{4d}\,S_{d}\Big).\nonumber 
\end{eqnarray}
This expression has a factor $\Lambda^{2d-8}(u_{\ast}^{(0)})^{2}$
appearing in both terms. So, as expected, expressing $u_{\ast}^{(0)}$
by \prettyref{eq:u_0_fixed} one obtains a universal result that does
not depend on $\Lambda$, the microscopic details of the system, reorganized
as a cubic equation of which we need to determine the roots

\begin{eqnarray}
0 & \simeq & \frac{8}{36}\,\frac{1}{S_{d}}\,\frac{2(3-d)-3\eta}{d}\,\Big(\frac{S_{d-1}}{2}\,(2-d-2\eta)\label{eq:anomalous_eta}\\
 &  & +\frac{S_{d}}{4}(3-d-2\eta)\,(4-d-3\eta)\Big)\,\big(4-d-2\eta\big)^{2}-\eta\,(3-d-2\eta)(2-d-2\eta).\nonumber 
\end{eqnarray}
For $3<d\le4$, we solve the cubic equation for $\eta$ numerically.
The result is shown in \prettyref{fig:Fixed-point-crit-exp}b.

Dropping from this cubic equation in $\eta$ all terms that are $\o(\eta^{2})$,
for $d=4$ we get a linear equation in $\eta$, relating it to ($u_{\ast}^{(0)})^{2}$
as
\begin{eqnarray}
\eta & \simeq & \frac{1}{16\pi^{8}}\,\frac{9S_{4}S_{3}}{2+\frac{1}{64\pi^{8}}\,(27S_{4}^{2}-90S_{4}S_{3})\,(u_{\ast}^{(0)})^{2}}\,(u_{\ast}^{(0)})^{2}\label{eq:eta_of_u0_4d}\\
 & \simeq & \frac{1}{16\pi^{8}}\,\frac{9S_{4}S_{3}}{2}\,(u_{\ast}^{(0)})^{2}+\o((u_{\ast}^{(0)})^{4})\nonumber \\
 & = & \frac{9}{4\pi^{5}}\,(u_{\ast}^{(0)})^{2}+\o((u_{\ast}^{(0)})^{4})\nonumber \\
 & \simeq & 0.0074\,(u_{\ast}^{(0)})^{2}+\o((u_{\ast}^{(0)})^{4}).\nonumber 
\end{eqnarray}
With $u_{\ast}^{(0)}\simeq\pi^{2}\,\frac{2}{9}\epsilon+\o(\epsilon^{2}$)
(cf. \prettyref{eq:u_0_3d_4d}) one sees that this result is consistent
with the limiting expressions for $d=4-\epsilon$ for $u_{\ast}^{(1)}\simeq\frac{8\pi}{27}\,\epsilon^{2}+\mathcal{O}(\epsilon^{3})$
(cf. \prettyref{eq:u1_fixed_3d_4d}) and $u_{\ast}^{(2)}=\frac{\pi^{2}}{72}\,\epsilon^{3}+\o(\epsilon^{4})$
inserted into \prettyref{eq:eta_base_eq} to find
\begin{eqnarray}
\eta & \simeq & \frac{1}{9\pi}\,\epsilon^{2}+\o(\epsilon^{3}).\label{eq:eta_d4}
\end{eqnarray}
The latter consideration shows that the $||k||$-dependence of the
interaction $u$ ($\propto u_{\ast}^{(1)}$) causes the anomalous
dimension close to four dimensions and the $k^{2}$-dependence of
the interaction ($\propto u_{\ast}^{(2)}$) only plays a role at higher
orders $\o(\epsilon^{3}$).

The value for $\eta$ \prettyref{eq:eta_d4} is by a factor $6/\pi\simeq1.9$
larger than the two-loop result from the $\epsilon$ expansion to
second order \cite[p. 626]{ZinnJustin96}
\begin{eqnarray}
\eta & \simeq & \frac{1}{54}\,\epsilon^{2}\label{eq:epsilon_expansion_second_order_eta}\\
 & \simeq & \frac{9}{24\,\pi^{4}}\,(u_{\ast}^{(0)})^{2}\nonumber \\
 & \simeq & 0.0038\,(u_{\ast}^{(0)})^{2}.\nonumber 
\end{eqnarray}

For $d=3$ we obtain a quadratic equation from \prettyref{eq:anomalous_eta}

\begin{eqnarray*}
0 & \simeq & \eta^{2}+\frac{17S_{3}+2S_{2}}{43S_{3}+4S_{2}}\,\eta-\frac{S_{2}}{43S_{3}+4S_{2}}\\
0 & \simeq & \eta^{2}+\frac{72}{180}\,\eta-\frac{1}{90},
\end{eqnarray*}
so
\begin{eqnarray}
\eta_{\pm}^{(d=3)} & = & -\frac{36}{180}\pm\sqrt{\big(\frac{36}{180}\big)^{2}+\frac{1}{90}}\label{eq:eta_3d}\\
 & = & \left\{ \begin{array}{cc}
0.0261 & +\\
-0.426 & -
\end{array}\right..\nonumber 
\end{eqnarray}
The positive solution is the physically relevant one.

\paragraph*{Critical exponent $\nu$}

The critical exponent $\nu$ is determined by linearizing the flow
about the fixed point and by determining the eigenvalue $\lambda_{r}$
in the direction of the parameter $r$. The critical exponent then
takes the value
\begin{eqnarray}
\nu & = & \lambda_{r}^{-1}\label{eq:nu_general}\\
 & = & \frac{1}{2-\eta-\frac{1}{18}\,(4-d-2\eta)(2(3-d)-3\eta)\,\big(\frac{6}{2(3-d)-3\eta}+\frac{4}{d-1}\,\frac{4-d-2\eta}{3-d-2\eta}\,\frac{S_{d-1}}{S_{d}}+\frac{1}{d}\,\frac{(4-d-3\eta)(4-d-2\eta)}{2-d-2\eta}\big)}.\nonumber 
\end{eqnarray}
So for $d=3$ and $d=4-\epsilon$ we obtain
\begin{eqnarray}
\nu & = & \left\{ \begin{array}{cc}
\frac{1}{2-\eta+\frac{1}{6}\,(1-2\eta)\big(-\frac{5}{2}+\eta-\eta\frac{1}{3}\,\frac{(1-3\eta)(1-2\eta)}{1+2\eta}\big)}\simeq\frac{1}{\frac{19}{12}-\frac{1}{18}\eta+\o(\eta^{2})} & d=3\\
\frac{1}{2-\frac{1}{3}\,\epsilon-\frac{8}{27\pi}\,\epsilon^{2}+\o(\epsilon^{3})} & d=4-\epsilon
\end{array}\right..\label{eq:nu_3d_4d}
\end{eqnarray}
For the approximation for $d=4-\epsilon$ we neglected $\eta\propto\epsilon^{2}$
throughout.

\section{Summary and discussion\label{sec:Summary-and-discussion}}

\subsection*{Summary of quantitative results for the $\varphi^{4}$-theory}

\begin{figure}
\begin{centering}
\includegraphics[width=0.9\textwidth]{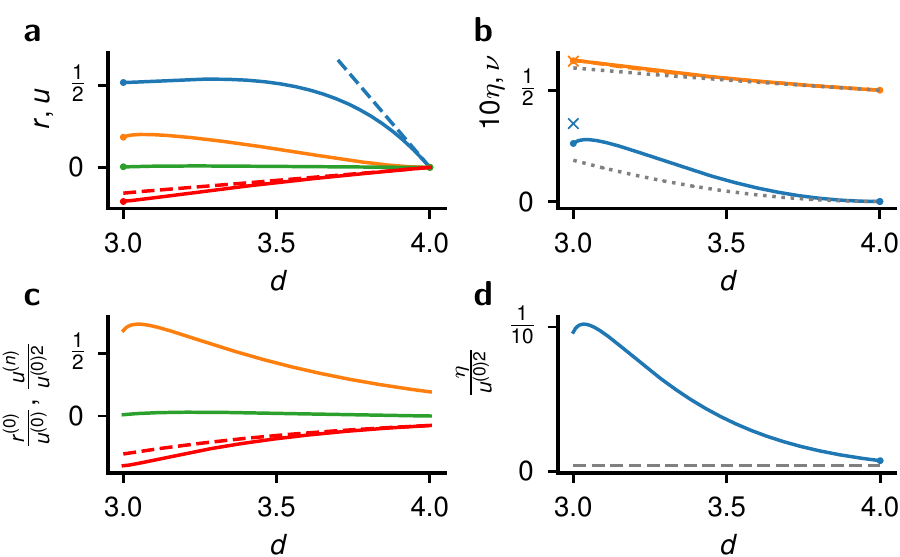}
\par\end{centering}
\caption{\textbf{Wilson-Fisher fixed point in $\varphi^{4}$-theory as function
of dimension $d\in[3,4]$. a} Fixed point as function of dimension
$d$: i) Mass $r_{\ast}^{(0)}$ (red solid curve, \prettyref{eq:r0_fixed};
red dashed curve: linear approximation in $\epsilon=4-d$; only first
line of \prettyref{eq:r0_fixed}) ii) Momentum-independent interaction
$u_{\ast}^{(0)}$ (blue curve, \prettyref{eq:u_0_fixed}; limits $d\to3,4$
\prettyref{eq:u_0_3d_4d} blue dot and blue dashed line). iii) First-order
momentum-dependence $u_{\ast}^{(1)}$ (orange curve \prettyref{eq:rel_u1}
and \prettyref{eq:u_0_fixed}; orange dot: limit $d\to3$ \prettyref{eq:u1_fixed_3d_4d}).
iv) Second-order momentum-dependence $u_{\ast}^{(2)}$ (green curve,
\prettyref{eq:rel_u2} and \prettyref{eq:u_0_fixed}; green dot: limit
$d\to3$ \prettyref{eq:u2_fixed_3d_4d}). \textbf{b} Critical exponents
$\nu$ and $\eta$ as function of dimension $d$: i) Anomalous dimension
$10\,\eta$ (blue curve, positive root of \prettyref{eq:anomalous_eta};
blue dots: limit $d\to3$ \prettyref{eq:eta_3d} and $d\to4$, $\eta=0$).
ii) Critical exponent $\nu$ (orange curve: \prettyref{eq:nu_general};
orange dots: limits $d\to3,4$ \prettyref{eq:nu_3d_4d}). Lowest order
approximations of $\epsilon$-expansion (dashed curves \prettyref{eq:lowest_order_eta}
for $\eta$ and $\nu=\big(2-(4-d)/3\big)^{-1}$ for $d=4-\epsilon$).
\textbf{c }Coupling constants relative to interaction: i) Relative
mass $r^{(0)}/u^{(0)}$ (red curve \prettyref{eq:rel_r0}; dashed
curve showing only linear dependent part on $u^{(0)}$, first line
in \prettyref{eq:rel_r0}). ii) Relative interaction $u^{(1)}/(u^{(0)})^{2}$
(orange curve \prettyref{eq:rel_u1})\textbf{ }and \textbf{$u^{(2)}/(u^{(0)})^{2}$}
(green curve \prettyref{eq:rel_u1}). Literature values of order $5$
$\epsilon$-expansion for $\eta$ (blue cross; $\eta=0.035$ \cite[p. 626]{ZinnJustin96}
and $\nu$ ($\nu=0.631(2)$ \cite[p. 635]{ZinnJustin96}).\textbf{
d }Relative anomalous dimension $\eta/(u^{(0)})^{2}$ (blue curve,
\prettyref{eq:rel_eta}; blue dot: limit for $d\to4$ \prettyref{eq:eta_of_u0_4d}
$\eta/(u_{\ast}^{(0)})^{2}=9/(4\pi^{5})\simeq0.0074$). Limit for
$d\to4$ of classical $\epsilon$ expansion to order $\epsilon^{2}$
(gray dashed line, \prettyref{eq:epsilon_expansion_second_order_eta}
$\eta/(u_{\ast}^{(0)})^{2}=9/(24\pi^{4})\simeq0.0038$).\label{fig:Fixed-point-crit-exp}}
\end{figure}

A summary of the derived expressions for the fixed point values of
the mass $r^{(0)}$, and the momentum-dependent interaction $\{u^{(0)},u^{(1)},u^{(2)}\}$
is shown in \prettyref{fig:Fixed-point-crit-exp}a for $d\in[3,4]$.

Critical exponents $\nu$ and $\eta$ as functions of the dimension
are shown in \prettyref{fig:Fixed-point-crit-exp}b. For $d=3$, the
obtained critical exponent $\nu\simeq0.632$ compares well to the
result of an order five $\epsilon$-expansion $\nu\simeq0.631(2)$
\cite[p. 635]{ZinnJustin96}. It is better that the result from the
momentum-scale expansion for the effective action ($\nu=0.532$) \cite{Morris96_477}
and to the second order $\epsilon$-expansion ($\nu=0.627$). The
increase in accuracy is despite the observation that the quadratic
momentum dependence $\o(k^{2})$ of the interaction vertex remains
small for all dimensions $d\in[3,4]$ (\prettyref{fig:Fixed-point-crit-exp}c).

Also the critical exponent $\eta\simeq0.026$ is not too far from
the result of the order five $\epsilon$-expansion of $\eta=0.035(3)$,
estimated from the divergent series in powers of $\epsilon$. The
here obtained value is closer to the true value than the two loop
result of $\eta\simeq1/54\simeq0.0185$ \cite[p. 625 eq. 28.7]{ZinnJustin96}.
This is so despite the simplicity of the computation presented here,
which only requires elementary one-loop integrals. Compared to the
momentum-scale expansion of the effective action \cite[p. 494]{Morris96_477}
($\eta=0.0225$), the improvement in accuracy is only small.

\begin{table}
\begin{centering}
\begin{tabular}{|c|c|c|c|c|}
\hline 
exponent & $\epsilon$-expansion $\o(\epsilon)$ & $\epsilon$-expansion $\o(\epsilon^{2})$ & this method & best approx.\tabularnewline
\hline 
\hline 
$\nu$ & $0.583$ & $0.627$ & $0.632$ & $0.631(2)$\tabularnewline
\hline 
$\eta$ & $0$ & $0.0185$ & $0.026$ & $0.035(3)$\tabularnewline
\hline 
\end{tabular}
\par\end{centering}
\caption{\textbf{Critical exponents in $d=3$ dimensions. }Comparison of this
method to known approximations. Best reference approximation from
\cite[p. 635]{ZinnJustin96} and \cite[p. 626]{ZinnJustin96}.}
\end{table}

The fixed point values in a Gell-Mann-Low theory, a theory with a
single marginal variable, can entirely be expressed in terms of this
single marginal variable \cite[Section V]{Wilson75_773}. This is
the basis of a renormalizable theory, where only a single renormalized
coupling constant must be fixed. Correspondingly, the resulting expressions
for the mass and for the momentum-dependent part of the interaction
can be expressed in terms of the momentum-independent part of the
interaction, $u^{(0)}$. Mass and the momentum-dependent part of the
interaction relative to $u^{(0)}$ (for the mass) or relative to $(u^{(0)})^{2}$
(for both interaction terms) are shown in \prettyref{fig:Fixed-point-crit-exp}c.

\subsection*{Relation to earlier work}

Even though the numerical values for the anomalous dimension $\eta$
obtained by the here-presented method and by an $\epsilon$-expansion
of second order are not so far apart, expressing the anomalous dimension
relative to $(u^{(0)})^{2}$ shows that the two methods are in fact
quite different: The $\epsilon$-expansion, to second order, yields
\cite[p. 625 eq. 28.7]{Wilson72_240,ZinnJustin96}
\begin{eqnarray}
\eta & = & \frac{1}{54}\epsilon^{2}+\o(\epsilon^{3})\label{eq:lowest_order_eta}
\end{eqnarray}
and thus, with the leading order behavior of $u^{(0)}=\pi^{2}\,\frac{2}{9}\epsilon+\o(\epsilon^{2})$
(\prettyref{fig:Fixed-point-crit-exp}a), a ratio independent of $\epsilon$
and thus $d$
\begin{eqnarray*}
\frac{\eta}{(u^{(0)})^{2}} & = & \frac{9}{24\,\pi^{4}}.
\end{eqnarray*}
The here-proposed method instead yields a saturating form of $u^{(0)}$
as the dimension is lowered from $d=4$ to $d=3$ (\prettyref{fig:Fixed-point-crit-exp}a),
while the relative quantity $\eta/(u^{(0)})^{2}$ increases as $d$
is lowered. Thus, the anomalous dimension, expressed in relation to
the renormalized coupling constant, is much larger in the here-proposed
method for $d=3$, as shown in \prettyref{fig:Fixed-point-crit-exp}d. 

\paragraph*{Relation to the work by Wegner \& Houghton \cite{Wegner73_401}}

The method proposed here is practically identical to the exact equations
derived by Wegner \& Houghton \cite{Wegner73_401}. The first main
contribution of the current manuscript is the presentation of an alternative
derivation without using the projector formulation and the discussion
how the momentum dependence of vertices leads to a tower of coupled
one-loop diagrams, that we show are of identical structure as the
known one-loop fluctuation expansion and the vertex expansion of the
Wetterich equation. The second contribution is to demonstrate that
this approach can in fact be combined with a momentum-scale expansion
to obtain quantitatively accurate results in a non-perturbative setting
far off the upper critical dimension.

It must be noted that the computation to order $\epsilon^{2}$ of
the anomalous dimension performed by Wegner and Houghton yields a
different result than found here. The cause of this difference is
the step from eq. (3.13) to eq. (3.16) in their work: Here the authors
approximately solve the set of their differential equations (3.9)
- (3.12). In doing so, they insert the contribution of the tree-diagrams
of the form \prettyref{eq:S6} contributing to the six-point vertex
$v_{6}$ given by their eq. (3.12) into the second line of their eq.
(3.11). What they thus implicitly assume is that the six-point vertex
$v_{6}$ is produced within the same decimation shell as the one integrated
out in the flow of $v_{4}$. This, however, is only true for vanishing
external momenta. Stated differently, they neglect that, for non-zero
external momenta $k$ of their four point vertex $v_{4}(k)$, the
two propagator lines in the composed diagram of the form \prettyref{eq:correction_four_point}
necessarily belong to different decimation shells. The propagator
line that connects the two four point vertices in their eq. (3.12)
of $v_{6}$, as a consequence, stems from a different decimation shell
than the propagator in their eq. (3.11). Our calculation takes this
momentum-dependence into account. This results in the factor $\left(\frac{\ell_{j}}{\ell_{i}}\right)^{6-3\eta-2d}$
in front of $S_{\ell_{i}}^{(6)}\big(\frac{\ell_{i}}{\ell_{j}}\,\{k\}\big)$
in \prettyref{eq:correction_four_point-2}, which is missing in Wegner
\& Houghton's approximate solution, their eq. (3.16). This additional
approximation they make yields a result for the critical exponent
$\eta$ that, to order $\epsilon^{2}$, is identical to the known
two-loop result. A careful interpretation of their equations in fact
yields the same result as presented in the current manuscript, had
they taken seriously the rescaling term $(6-3\eta-2d)\,v_{6}$ in
their eq. (3.12) in conjunction with the momentum dependence, as explained
in \prettyref{fig:Decimation-causes-flow-momentum}.

A closely related approach is the functional renormalization group
\cite{WETTERICH93_90,Morris94_2411} for the effective action, the
first Legendre transform of the free energy. Employing a hard cutoff,
this method has been used previously to compute the momentum dependence
of the self-energy in weakly interacting Bosons \cite{Hasselmann04_101103,Ledowski04_101103}.
Here as well, a momentum-scale expansion of the interaction leads
to a non-differentiable momentum dependence $\propto||k||$ of the
interaction vertex that enters the computation of the anomalous dimension
(cf. i.p. \cite{Ledowski04_101103} eq. (11)) and that becomes marginal
in $d=3$ dimensions, as in the present manuscript (cf. eq. \prettyref{eq:flow_eq_u1}).

\paragraph*{Relation to the $\epsilon$-expansion}

An important cross check of the method is its compatibility with
previous methods, foremost the established $\epsilon$-expansion \cite{Wilson72_240,Wilson74_75,Wilson75_773}.
The interaction $u_{\ast}^{(0)}\simeq\pi^{2}\,\frac{2}{9}\epsilon+\o(\epsilon^{2})$,
given by \prettyref{eq:u_0_3d_4d}, is in fact identical to order
$\o(\epsilon)$ in the two approaches. Also, the momentum dependence
only arises at order $\o(\epsilon^{2})$, given by \prettyref{eq:u1_fixed_3d_4d}
and \prettyref{eq:u2_fixed_3d_4d}, again in line with the $\epsilon$-expansion.
A qualitative difference is, though, that the leading order $\o(\epsilon^{2})$
momentum-dependence of the interaction vertex here is $\propto u_{\ast}^{(1)}\,||k_{1}+k_{2}||$
with $\Lambda^{d-3}u_{\ast}^{(1)}\simeq\frac{8\pi}{27}\,\epsilon^{2}+\mathcal{O}(\epsilon^{3})$,
whereas the quadratic momentum-dependence $\propto u_{\ast}^{(2)}\,(k_{1}+k_{2})^{2}$
only arises at the next order in $\epsilon$, namely $\Lambda^{d-2}u_{\ast}^{(2)}\simeq\frac{\pi^{2}}{72}\,\epsilon^{3}+\o(\epsilon^{4})$.
The correction to $\eta$ to order $\o(\epsilon^{2})$ comes from
the third line of the self-energy correction \prettyref{eq:correction_self_energy_infinitesimal_anomalous-1}
by inserting the momentum-dependence $\propto u_{\ast}^{(1)}$ in
the middle term of \prettyref{eq:q_squared_S4}.

Such contribution $\propto u_{\ast}^{(1)}$ is not present in the
classical second order $\epsilon$-expansion, because this method
treats the momentum dependence of the interaction differently. The
leading order $k$-dependence of the self-energy in the $\epsilon$-expansion
comes from the diagram (cf. \cite[Fig 5.4]{Wilson74_75})
\begin{eqnarray}
 &  & \Diagram{\\
\vertexlabel_{k}g & fflfluf & g\vertexlabel_{-k}.
}
\label{eq:two_loop_self_energy}\\
\nonumber 
\end{eqnarray}
The $k$-dependence of this diagram results from an expansion in the
external momentum $k$. This diagram can also be seen as arising from
contracting a pair of legs of the one-loop fluctuation correction
to the four point vertex, $\Gamma_{\mathrm{fl}}^{(4)}$, as in eq.
\prettyref{eq:correction_self_energy_infinitesimal_anomalous-1}.
This sub-diagram of \prettyref{eq:two_loop_self_energy} is of the
form
\begin{eqnarray}
-\Gamma_{\mathrm{fl}}^{(4)}(q,k,-k,-q) & = & \Diagram{\\
\vertexlabel^{q}\;gd & f0flfluf0 & gu\,\vertexlabel^{-q}\\
\vertexlabel^{k}\;gu &  & gd\,\vertexlabel^{-k}
}
\quad,\label{eq:Gamma_fl_4}\\
\nonumber 
\end{eqnarray}
where $q$ is the momentum of integration of the upper loop in \prettyref{eq:two_loop_self_energy}.
The dependence on $k$ then results from the algebraic form of the
diagram
\begin{eqnarray}
-\Gamma_{\mathrm{fl}}^{(4)}(q,k,-k,-q) & \propto & \int\frac{dk^{\prime}}{(2\pi)^{d}}\,\frac{u_{0}^{2}}{(r_{0}+(q+k^{\prime})^{2})\,(r_{0}+(k-k^{\prime})^{2})},\label{eq:fluct_corr_Gamma4}
\end{eqnarray}
when expanding to quadratic order in $k$: One obtains only spherically
symmetric diagrams $\propto k^{2}$ and uniaxial diagrams $\int dk^{\prime}\,(k\cdot k^{\prime})^{2}\propto k^{2}$;
to lowest order in $k$ one thus obtains a dependence $\propto k^{2}$.
In particular, there is no dependence $\propto||k||$ arising in the
perturbative approach, because expanding for small $k$, terms $\propto\int dk^{\prime}\,(k\cdot k^{\prime})$
vanish by the point-symmetry of the integrand. In contrast, in our
approach we have found this very term $\propto u_{\ast}^{(1)}||k||$
to produce the dominant contribution to $\eta$.

What is the reason for this qualitative difference between the two
approaches? The non-differentiable momentum dependence $\propto u_{\ast}^{(1)}||k||$
of the interaction in our approach comes about by the fact that a
fluctuation correction of the form \prettyref{eq:Gamma_fl_4} at scale
$\ell$ necessarily must have one of the propagator lines above the
current cutoff $|q+k^{\prime}|>\Lambda_{\ell}$ or $|k-k^{\prime}|>\Lambda_{\ell}$.
When expanded for small momenta, this leads to integrals of the form
$\propto\int dk^{\prime}\,(k\cdot k^{\prime})\,\theta(k\cdot k^{\prime})\propto||k||$,
where $\theta(k\cdot k^{\prime})$ constrains the integration to half
the momentum shell. These contributions thus do not vanish, as opposed
to the perturbative approach, where one integrates over full momentum
shells.

On a more abstract level, we can summarize these observations as follows.
The infinitesimal Wilsonian RG studied here leads to a functional
differential equation of the form
\begin{eqnarray}
\ell\frac{dS_{\ell}}{d\ell} & = & \beta\{S_{\ell}\},\label{eq:func_diffeq}
\end{eqnarray}
where $\beta:\mathcal{F}\mapsto\mathcal{F}$ is a mapping from the
space of functionals $\mathcal{F}:=\{f|f:\mathcal{C}\mapsto\mathbb{R}\}$
into itself, where $\mathcal{C}$ is the space of functions, in our
case the configurations $\varphi$ of our system. In the chosen Fourier
representation of $\varphi$, the functional mapping $\beta$ may
thus couple different $k$-vectors. This is in fact the case, as we
illustrate in \prettyref{fig:Decimation-causes-flow-momentum}: The
flow for the Taylor coefficient $S_{\ell}^{(4)}(q_{1},q_{2},q_{1}^{\prime},q_{2}^{\prime})$
of $S_{\ell}$ depends on the Taylor coefficient $S_{\ell}^{(6)}(q_{1},q_{2},q_{3},\ldots)$;
here $S_{\ell}^{(4)}$ appears on the left hand side of \prettyref{eq:func_diffeq}
and $S_{\ell}^{(6)}$ on the right hand side. The here-followed approach
takes this functional seriously; as a result, the value $S_{\ell}^{(6)}$
that appears on the right hand side of \prettyref{eq:func_diffeq}
depends on the momentum $q_{1}+q_{2}+q_{3}$ given by the arguments
$q_{1}$ and $q_{2}$ of $S_{\ell}^{(4)}$ and the loop momentum $q_{3}$,
because the decimation step $\ell_{i}(q_{1}+q_{2}+q_{3})$ in which
$S_{\ell}^{(6)}$ has been produced is a function of this very momentum,
as explained in \prettyref{sec:Momentum-dependence}.

This is what appears to us being neglected in the perturbative approach
of the $\epsilon$-expansion: Here, the diagram \prettyref{eq:fluct_corr_Gamma4}
is composed of vertices $u_{0}$ and propagators $(r_{0}+k^{2})^{-1}$,
all of which take the values of the current decimation step; the perturbative
calculation thus neglects the fact that each propagator line necessarily
belongs to a single decimation step that integrates out the degrees
of freedom at the very momentum that is carried by the propagator.

The difference between the perturbative calculation and our approach
consequently shows up only in the flow of vertices at non-zero external
momenta; the value for $u_{0}$ at vanishing momentum to order $\epsilon$
is identical. In contrast, the two methods yield quantitatively different
results for the anomalous dimension, because this quantity is defined
in terms of the momentum dependence of the self-energy.

The estimate of the anomalous dimension by the $\epsilon$-expansion,
to quadratic order $\o(\epsilon^{2})$, is roughly by about a factor
$2$ smaller than the estimate of a high temperature expansion, as
already noted in Fig. 1 of \cite{Wilson72_240}. In their ref 16,
the authors point out that their estimate is in fact too small. Also
the comment after eq. (8.29) on p. 137 of \cite{Wilson74_75} points
out that the result of the $\epsilon$-expansion at second order is
too small by about a factor of $2$. Our estimate is larger by a factor
$6/\pi\simeq1.9$ than the $\epsilon$-expansion at second order.
The here-presented method is thus in better agreement with the high
temperature expansion and with higher orders of the $\epsilon$-expansion.
However, the approximations made here, truncating the vertex expansion
and expanding in small momenta, requires further careful analysis
to see if this improvement is truly systematic.

On the technical side a difference of the proposed method is that
it does not require resummation of a divergent series, as opposed
to the $\epsilon$-expansion. The flow equations expose the desired
quantities simply by their fixed point values. This shows the non-perturbative
nature of the method. The approximation made here is not in the number
of loops \textendash{} we have shown above that the one-loop calculation
is exact; this is simply a result of the phase space volume contributing
to an infinitesimal marginalization step being proportional to the
(infinitesimal) thickness of the momentum shell times the number of
loops. Formally, this result is thus in line with exact functional
RG equations \cite{Wegner73_401,WETTERICH93_90,Morris94_2411}, where
also only single-loop integrals are required. For the functional RG
of the effective action \cite{WETTERICH93_90,Morris94_2411}, similar
approximations, a momentum scale expansion combined with the vertex
expansion, have been applied to investigate critical behavior of weakly
interacting Bosons and in particular to obtain an approximation of
the the momentum dependence of the self-energy and thus the anomalous
dimension \cite{Hasselmann04_101103,Ledowski04_101103}.  

In the simple example shown here we did not compute the flow of the
six-point vertex. At $d=3$, however, the six point vertex becomes
marginal so that its contribution can potentially become substantial.
Since the $|k|$-momentum-dependence of the six point vertex is less
relevant by one power of $\ell$, an approximation taking the the
momentum-independent part into account is still simple and yields
potentially better results. Also, taking into account the momentum
dependence of the four-point vertex in computing its own corrections
is expected to improve the results close to $d=3$.

\subsection{Conclusion and outlook}

We here analyzed the exact Wegner-Houghton renormalization group equation,
here derived as an infinitesimal momentum-shell Wilsonian RG. A vertex
expansion exposes the close relation to the Wetterich RG with hard
cutoff. We found that the flow equations can successfully be closed
by combining the vertex expansion with a momentum-scale approximation.
For the scalar $\varphi^{4}$-theory, one obtains a highly accurate
approximation for $\nu$ and a reasonable accurate approximation for
$\eta$; even though, the latter vanishes at one-loop order of the
orthodox $\epsilon$-expansion.

Our main motivation is to render a more complete picture on the relation
between the Wegner-Houghton RG and the Wilson RG and $\epsilon$-expansion
as well as the Wetterich RG; in particular, to point out where approximations
are being made. The main result is a tower of coupled flow equations
for the momentum-dependent vertices that is logically consistent with
a continuous marginalization of modes.

The obvious weakness of the here-presented solution of this set of
equations is the requirement of truncating the vertex expansion and
the treatment of the momentum dependence; in both points one faces
basically the same problem as in the vertex expansion of the Wetterich
RG. Even though resummation techniques in $\epsilon=d_{c}-d$ are
not required here (neither in the Wetterich RG), which may a priori
be regarded as an advantage, there is of course a large body of literature
that has carefully studied the behavior of these asymptotic series.
More work is needed to thoroughly study the limits of the vertex expansion
combined with the momentum-scale approximation used here.

The use of the renormalization group to study phenomena that arise
from the interaction of processes on many different scales is not
restricted to systems from the core domains of physics. Our motivation
to employ the renormalization group, for example, arises from the
wish to understand how activity organizes in neuronal systems. Field
theoretical formulations of neuronal networks have begun to be employed
in this field (see e.g. \cite{Chow15,Hertz16_033001,Helias19_10416}
for reviews). One here knows the dynamical equations on the microscopic
level, namely for individual neurons, but experiments often only yield
coarse-grained signals, such as the local-field potential \cite{Linden11_859}.
Moreover, there is experimental evidence that neuronal systems may
operate at critical points \cite{Beggs03_11167}. Methods from statistical
physics are therefore commonly employed in this field; often from
equilibrium statistical mechanics. For example the pairwise maximum-entropy
model \cite{Roudi09}, the Ising model, is fitted to binned neuronal
activity data \cite{Mora2011}. More recent developments have formulated
neuronal systems as a Ginzburg-Landau field theory, a genuine non-equilibrium
formulation \cite{diSanto18_1356}. Such systems bear strong similarity
to the Kardar-Parisi-Zhang model \cite{Kardar84}. For the latter
model it is known that the momentum dependence of the self-energy
is a key ingredient of the flow equations. Such models thus necessitate
efficient approximations of the momentum dependence, as investigated
here. But not only the study of coarse-graining and critical dynamics
in biological neuronal networks requires renormalization group techniques.
Also the transformations performed by deep artificial neuronal networks
are composed of concatenations of large numbers of relatively simple
transformations by each layer. Addressing the question how representations
of data arise in these networks \cite{Bengio13_1798} may therefore
also be studied by renormalization group methods \cite{Mehta14renormalization}.
Applying an established technique to a new field, often shows a method
in a new light. This is how the question of the momentum dependence
in the Wilson RG, as discussed by the current manuscript, arose in
the present case.

We hope that the here presented method may become fruitful to obtain
simple and intuitive diagrammatic approximations for critical phenomena
in various systems that so far required more elaborate non-perturbative
methods. We also hope that the partly pedagogical presentation may
be helpful for the accessibility by a broad readership.

\section{Acknowledgments}

We are grateful to the comments by Peter Kopietz on an earlier version
of this manuscript. These comments in particular considerably improved
the discussion of the presented work to the $\epsilon$-expansion
and to non-perturbative techniques using the functional RG and the
momentum-scale expansion. We also thank the anonymous reviewer for
the helpful and constructive critique and for pointing out the common
misunderstanding that the field-theoretical RG is only applicable
to renormalizable field theories. We would further like to acknowledge
helpful discussions with Tobias Kuehn, and Kirsten Fischer that lead
to the initial question on the momentum-dependence of the infinitesimal
RG.

This work was partially supported by the European Union's Horizon
2020 research and innovation programme under grant agreement No.\ 785907
(Human Brain Project SGA2), the Exploratory Research Space (ERS) seed
fund G:(DE-82)EXS-SF-neuroIC002, and the German Federal Ministry for
Education and Research (BMBF Grant 01IS19077A).

\section{Appendix}

\subsection{Gaussian part of the action\label{sec:Gaussian_model}}

Before proceeding to the general problem to apply the coarse-graining
defined in \prettyref{sec:Definition-of-coarse-graining} to the full
action, consider the quadratic term which is
\begin{eqnarray}
S_{0}[\varphi]:=-\frac{1}{2}\int_{k}\,\varphi(-k)\,(r^{(0)}+r^{(2)}\,k^{2})\,\varphi(k) & = & :-\frac{1}{2}\varphi^{\T}G^{-1}\varphi,\label{eq:Gaussian_action_Wilson}
\end{eqnarray}
where we introduce a matrix-vector notation and define

\begin{eqnarray*}
G^{-1}(k,k^{\prime}) & = & \delta(k+k^{\prime})\,(r^{(0)}+r^{(2)}\,k^{2}).
\end{eqnarray*}
This matrix is diagonal in the sense that it only couples the component
$\varphi(k)$ with $\varphi(-k)$.

To perform the marginalization step \prettyref{eq:coarse_grained_action}
one needs to evaluate $S_{0}[\varphi_{<}+\varphi_{>}]$
\begin{eqnarray}
\varphi^{\T}G^{-1}\varphi & = & (\varphi_{<}+\varphi_{>})^{\T}\,G^{-1}\,(\varphi_{<}+\varphi_{>}).\label{eq:square_part_Wilson}
\end{eqnarray}
As $G^{-1}$ is diagonal in frequency domain, it couples only frequencies
$0<|q|<\Lambda\ell^{-1}$ among another and frequencies $\Lambda_{\ell}:=\Lambda\ell^{-1}<|k|<\Lambda$
among another, but it does not couple $q$ and $k$; cross terms in
multiplying out \prettyref{eq:square_part_Wilson} are thus dropped
to get
\begin{eqnarray}
\varphi^{\T}G^{-1}\varphi & = & \varphi_{<}^{\T}\,G^{-1}\varphi_{<}\label{eq:quadratic_Wilson_expanded}\\
 & + & \varphi_{>}^{\T}\,G^{-1}\varphi_{>}.\nonumber 
\end{eqnarray}
The decimation step in \prettyref{eq:coarse_grained_action} thus
only affects the term in the second line of \prettyref{eq:quadratic_Wilson_expanded},
which yields
\begin{eqnarray}
\Z_{>} & := & \int\D\varphi_{>}\,\exp(-\frac{1}{2}\,\varphi_{>}^{\T}\,G^{-1}\varphi_{>})\label{eq:Z_larger}\\
 & = & \prod_{\Lambda_{\ell}<|k|<\Lambda}\,\big(2\pi\,G(k)\big){}^{\frac{1}{2}}\nonumber \\
\ln\Z_{>} & = & \frac{1}{2}\int_{\Lambda_{\ell}<|k|<\Lambda}\,\ln2\pi G(k).\nonumber 
\end{eqnarray}
This contribution is hence a multiplicative factor $\Z_{>}$ changing
the partition function, independent of $\varphi_{<}$. Thus it affects
the free energy. But it has no consequence on the correlation functions
of the field, because it becomes an additive correction to the cumulant-generating
functional $W=\ln\Z$. Since we here want to calculate only the critical
exponents, we will neglect this term. 

The remaining part of the coarse-grained action stems from the first
line in \prettyref{eq:quadratic_Wilson_expanded}, which just yields
\begin{eqnarray*}
 & \exp & (-\frac{1}{2}\varphi_{<}^{\T}\,G^{-1}\varphi_{<}).
\end{eqnarray*}
So together the coarse-grained action is
\begin{eqnarray}
S_{\ell}[\varphi_{<}] & = & \ln\Big(\,\Z_{>}\,\exp(-\frac{1}{2}\varphi_{<}^{\T}\,G^{-1}\varphi_{<})\Big)\label{eq:coarse_grained_Gaussian}\\
 & = & \ln\,\Z_{>}-\frac{1}{2}\varphi_{<}^{\T}\,G^{-1}\varphi_{<}.\nonumber 
\end{eqnarray}
The last line shows that the rescaled action, apart from the inconsequential
constant, is the same as the original one \prettyref{eq:Gaussian_action_Wilson}.
With one exception: the momenta of the fields only extend up to $\Lambda\ell^{-1}$
instead of $\Lambda$.

\subsection{Expansion of momentum dependence of rescaling term\label{sec:Expansion-of-momentum-dependence}}

The momentum dependence of the rescaling term appearing in \prettyref{eq:correction_four_point-2},
to order $\text{\ensuremath{\o}}(k_{\ell}^{2})$, takes the form

\begin{eqnarray}
\left(\frac{\ell_{j}}{\ell_{i}}\right)^{6\,\big(1-\frac{d+\eta}{2}\big)+d} & = & \exp\Bigg(\Big[6\,\big(1-\frac{d+\eta}{2}\big)+d\Big]\,\frac{1}{2}\ln\Big(\frac{\ell_{j}}{\ell_{i}}\Big)^{2}\Bigg)\label{eq:mom_rescaling}\\
 & = & \exp\Bigg(\Big[6\,\big(1-\frac{d+\eta}{2}\big)+d\Big]\,\frac{1}{2}\ln\Big(\frac{(k^{\prime}+k)^{2}}{\Lambda^{2}}\Big)\Bigg)\nonumber \\
 & = & \exp\Bigg(\Big[6\,\big(1-\frac{d+\eta}{2}\big)+d\Big]\,\frac{1}{2}\ln\Big(\frac{\overbrace{k^{\prime2}}^{\Lambda^{2}}+2\,k^{\prime}\cdot k+k^{2}}{\Lambda^{2}}\Big)\Bigg)\nonumber \\
 & = & \exp\Bigg(\Big[6\,\big(1-\frac{d+\eta}{2}\big)+d\Big]\,\frac{1}{2}\ln\Big(1+\frac{2\,k^{\prime}\cdot k+k^{2}}{\Lambda^{2}}\Big)\Bigg)\nonumber \\
 & \simeq & \exp\Bigg(\Big[6\,\big(1-\frac{d+\eta}{2}\big)+d\Big]\,\Big(\frac{1}{2}\,\frac{2\,k^{\prime}\cdot k+k^{2}}{\Lambda^{2}}-\frac{1}{2}\,\frac{1}{2}\,\frac{4\,(k^{\prime}\cdot k)^{2}}{\Lambda^{4}}\Big)+\mathcal{O}(k^{3})\Bigg)\nonumber \\
 & \simeq & 1+\Big[6\,\big(1-\frac{d+\eta}{2}\big)+d\Big]\,\Big(\frac{2\,k^{\prime}\cdot k+k^{2}}{2\,\Lambda^{2}}-\frac{(k^{\prime}\cdot k)^{2}}{\Lambda^{4}}\Big)\nonumber \\
 & + & \frac{1}{2!}\,\Big[6\,\big(1-\frac{d+\eta}{2}\big)+d\Big]^{2}\,\frac{(k^{\prime}\cdot k)^{2}}{\Lambda^{4}}+\mathcal{O}(k^{3})\nonumber \\
 & = & 1\nonumber \\
 & + & \frac{(k^{\prime}\cdot k)}{\Lambda^{2}}\,\Big(6\,\big(1-\frac{d+\eta}{2}\big)+d\Big)\nonumber \\
 & + & \frac{k^{2}}{2\,\Lambda^{2}}\,\Big(6\,\big(1-\frac{d+\eta}{2}\big)+d\Big)\nonumber \\
 & + & \frac{(k^{\prime}\cdot k)^{2}}{\Lambda^{4}}\,\Big[\frac{1}{2}\,\Big(6\,\big(1-\frac{d+\eta}{2}\big)+d\Big)^{2}-\Big(6\,\big(1-\frac{d+\eta}{2}\big)+d\Big)\Big]\nonumber \\
 & = & 1+\Big(6-2\,d-3\eta\Big)\,\Bigg(\frac{(k^{\prime}\cdot k)}{\Lambda^{2}}+\frac{k^{2}}{2\,\Lambda^{2}}+\frac{(k^{\prime}\cdot k)^{2}}{\Lambda^{4}}\,\Big[2-d-\frac{3}{2}\eta\Big]\Bigg).\nonumber 
\end{eqnarray}

\subsection{Uniaxial and spherical symmetric integrals\label{sub:Uniaxial-spherical-symmetric}}

We write for short
\begin{eqnarray*}
\int d\Omega\,f(\Omega\cdot\Lambda) & \equiv & \int_{|k|=\Lambda}f(k).
\end{eqnarray*}
If the integrand is spherically symmetric, so that it does not depend
on $\Omega$, we may perform the angular integral to obtain
\begin{eqnarray}
\int d\Omega & = & S_{d}=\frac{2\pi^{\frac{d}{2}}}{\Gamma(\frac{d}{2})},\label{eq:def_S_d}
\end{eqnarray}
which yields the area of the $d$-dimensional unit sphere and $\Gamma$
denotes the gamma function $\Gamma(z)=\int_{0}^{\infty}t^{z-1}\,e^{-t}\,dt.$

For uniaxial integrals we have \cite[p. 379]{Goldenfeld92}
\begin{eqnarray}
I & = & \int\frac{d^{d}q}{(2\pi)^{d}}\,f(|q|,\theta)\label{eq:uniaxial}\\
 & = & \int dq\,q^{d-1}\int d\Omega\,f(|q|,\theta)\nonumber \\
 & = & \int dq\,q^{d-1}C\,\int_{0}^{\pi}d\theta\,(\sin\theta)^{d-2}\,f(|q|,\theta),\nonumber \\
C & = & \frac{S_{d}}{(2\pi)^{d}}\,\frac{1}{\int_{0}^{\pi}d\theta\,(\sin\theta)^{d-2}}=\frac{S_{d-1}}{(2\pi)^{d}}.\nonumber 
\end{eqnarray}
We need such integrals to compute the $k$-dependence of the fluctuation
corrections. Here the integrand $f(|q|,\theta)$ only depends on the
absolute magnitude $|q|$ and a single angle $\theta$ between a specified
direction in $\mathbb{R}^{d}$ and the $q$-vector to integrate over.

We need in particular integrands of the form $f(|q|)\,\cos^{2}(\theta)$
for which it holds that

\begin{eqnarray}
J & = & \int\frac{d^{d}q}{(2\pi)^{d}}\,f(|q|)\,\cos^{2}(\theta)\label{eq:J_uniaxial}\\
 & = & \frac{1}{(2\pi)^{d}}\,\int d\Omega\,\cos^{2}(\theta)\,\int dq\,q^{d-1}\,f(|q\text{|)}\nonumber \\
 & = & \frac{S_{d}}{(2\pi)^{d}}\,\frac{1}{d}\,\int dq\,q^{d-1}\,f(|q\text{|)}.\nonumber 
\end{eqnarray}
This can be seen as follows: For the angular part of \prettyref{eq:J_uniaxial}
alone we get with \prettyref{eq:uniaxial}
\begin{eqnarray}
\int d\Omega\,\cos^{2}(\theta) & = & C\,\int_{0}^{\pi}d\theta\,\sin^{d-2}(\theta)\,\cos^{2}(\theta)\label{eq:angular_int-1-1}\\
 & = & C\,\int_{0}^{\pi}d\theta\,\sin^{d-2}(\theta)\,(1-\sin^{2}(\theta))\nonumber \\
 & = & C\,\int_{0}^{\pi}d\theta\,\sin^{d-2}(\theta)-\sin^{d}(\theta).\nonumber 
\end{eqnarray}
We may reduce the latter integral to the former for $d>1$
\begin{eqnarray*}
I:=\int_{0}^{\pi}d\theta\,(\sin\theta)^{d} & = & \int_{0}^{\pi}d\theta\,\underbrace{\sin\theta}_{f^{\prime}}\underbrace{(\sin\theta)^{d-1}}_{g}\\
 & = & \underbrace{-\cos\theta\,(\sin\theta)^{d-1}\Big|_{0}^{\pi}}_{0}+\int_{0}^{\pi}d\theta\,\cos^{2}\theta\,(d-1)\,(\sin\theta)^{d-2}\\
 & = & (d-1)\,\int_{0}^{\pi}d\theta\,(1-\sin^{2}\theta)\,(\sin\theta)^{d-2}\\
 & = & (d-1)\,\int_{0}^{\pi}d\theta\,(\sin\theta)^{d-2}-(d-1)\,I.
\end{eqnarray*}
So we get
\begin{eqnarray*}
I & = & \frac{d-1}{d}\,\int_{0}^{\pi}d\theta\,(\sin\theta)^{d-2}.
\end{eqnarray*}
We can therefore combine both angular integrals \prettyref{eq:angular_int-1-1}
as
\begin{eqnarray}
\frac{1}{(2\pi)^{d}}\,\int d\Omega\,\cos^{2}(\theta) & = & \frac{S_{d}}{(2\pi)^{d}}\,\big(1-\frac{d-1}{d}\big)=\frac{S_{d}}{(2\pi)^{d}}\,\frac{1}{d},\label{eq:int_cos_squared-1}
\end{eqnarray}
from which follows \prettyref{eq:J_uniaxial}.

\subsection{Momentum-dependent integral of four-point vertex\label{sec:Momentum-dependent-integrals-of-four-point}}

Here we compute the integral appearing in \prettyref{eq:correction_four_point-2}
at small momenta $k$, using \prettyref{eq:mom_rescaling}. We hence
have to compute two spherically symmetric and two uniaxial terms.
The latter uniaxial term $\propto(k^{\prime}\cdot k)^{2}$ in \prettyref{eq:mom_rescaling}
yields with $(k\cdot k^{\prime})^{2}=||k||^{2}\,\Lambda^{2}\,\cos(\Omega_{1})^{2}$
and \prettyref{eq:int_cos_squared-1}
\begin{eqnarray}
\int_{|k^{\prime}|=\Lambda}\,\theta(k^{\prime}\cdot k+)\,(k\cdot k^{\prime})^{2} & = & \frac{1}{2}\,\Lambda^{2}\,k^{2}\,\frac{S_{d}}{d},\label{eq:uniaxial_1}
\end{eqnarray}
where the factor $1/2$ comes from the constraint that by $\theta(k^{\prime}\cdot k_{\ell})$
we are only integrating over half a momentum shell. The spherically
symmetric integrals ($\propto(k^{\prime})^{0}$) simply produce one
factor $S_{d}$; here the entire shell contributes for vanishing external
momenta. The remaining integral is given with \prettyref{eq:uniaxial}
by
\begin{eqnarray}
 &  & \int_{|k^{\prime}|=\Lambda}\,\theta(k^{\prime}\cdot k+)\,k\cdot k^{\prime}\label{eq:uniaxial_2}\\
 & = & \Lambda||k||\,\int_{|k^{\prime}|=\Lambda}\,\theta(\cos(\Omega_{1})+)\,\cos(\Omega_{1})\nonumber \\
 & \stackrel{(\ref{eq:uniaxial})}{=} & \Lambda S_{d-1}\,||k||\,\int_{0}^{\pi}d\Omega_{1}\,(\sin\Omega_{1})^{d-2}\,\theta(\cos(\Omega_{1})+)\,\cos(\Omega_{1})\nonumber \\
 & = & S_{d-1}\,\Lambda||k||\,\int_{0}^{\pi/2}d\Omega_{1}\,(\sin\Omega_{1})^{d-2}\,\cos(\Omega_{1})\nonumber \\
 & = & S_{d-1}\,\Lambda||k||\,\frac{1}{d-1}\,(\sin\Omega_{1})^{d-1}\big|_{0}^{\pi/2}=\Lambda S_{d-1}\,||k||\,\frac{1}{d-1}.\nonumber 
\end{eqnarray}
Taken together, we get \prettyref{eq:decimation_S4_non_approx} in
the main text.

\subsection{Momentum dependence of self-energy\label{sec:Momentum-dependence-of-self-energy}}

Approximating the dependence on the momentum $q$ for $q\ll k$ up
to orders of $\o(q^{2})$, the resulting integrals appearing in \prettyref{eq:correction_self_energy_infinitesimal_anomalous-1}
are again either spherically symmetric or uniaxial which follows with
$\sqrt{1+x}\simeq1+\frac{x}{2}-\frac{x^{2}}{8}+\text{\ensuremath{\o}}(x^{2})$
from the expansion \prettyref{eq:S4_momentum_dep}
\begin{eqnarray*}
 &  & \frac{1}{4!}S_{\ell}^{(4)}(q,k,-k,q)\\
 & \simeq & u_{\ell}^{(0)}\\
 & + & u_{\ell}^{(1)}\,||k||\,\Big(1+\underbrace{\frac{q\cdot k}{||k||^{2}}}_{\to0}+\frac{1}{2}\,\frac{q^{2}}{||k||^{2}}-\frac{1}{2}\,\frac{(q\cdot k)^{2}}{||k||^{4}}\Big)\\
 & + & u_{\ell}^{(2)}\,(q^{2}+\underbrace{2q\cdot k}_{\to0}+k^{2})+\o(\frac{||q||^{3}}{||k||^{3}}).
\end{eqnarray*}
Taking the integral over the sphere of $k$-vectors in \prettyref{eq:correction_self_energy_infinitesimal_anomalous-1},
the terms $\propto2q\cdot k$ do not contribute by their point symmetry,
indicated by the underbraces. The terms $\propto q^{2}$ yield a momentum
dependence of the self-energy. The uniaxial term is again of the form
\prettyref{eq:J_uniaxial} and yields
\begin{eqnarray*}
-\frac{u_{\ell}^{(1)}}{2}\,\int_{|k|=\Lambda_{\ell}}\,\frac{(q\cdot k)^{2}}{||k||^{3}} & = & -\frac{u_{\ell}^{(1)}}{2\Lambda_{\ell}}\,\frac{S_{d}}{d}\,q^{2}.
\end{eqnarray*}
To determine the flow equations for $r^{(0)}$ and $r^{(2)}$, we
need to decompose the self-energy corrections $\ell\frac{d}{d\ell}S_{\ell}^{(2)}(q,-q)$
according to their $q^{2}$-dependence as
\begin{eqnarray}
\ell\frac{d}{d\ell}r_{\ell}^{(0)} & = & -\ell\frac{d}{d\ell}\,S_{\ell}^{(2)}(0,0),\label{eq:projection_self_energy}\\
\ell\frac{d}{d\ell}r_{\ell}^{(2)} & = & -\lim_{q\to0}\,\frac{\ell\frac{d}{d\ell}S_{\ell}^{(2)}(q,-q)-\ell\frac{d}{d\ell}S_{\ell}^{(2)}(0,0)}{q^{2}}.\nonumber 
\end{eqnarray}
It is therefore sufficient to compute \prettyref{eq:correction_self_energy_infinitesimal_anomalous-1}
for small momenta $q$. This, in turn, means we need the $q$-dependence
of the interaction vertex in the first two entries, given by \prettyref{eq:S4_momentum_dep}.
The projections \prettyref{eq:projection_self_energy} hence yield
the flow equation for the momentum-independent parameter
\begin{eqnarray*}
\ell\frac{d}{d\ell}r_{\ell}^{(0)} & = & 2\cdot2\cdot\frac{\Lambda^{d}}{(2\pi)^{d}}\cdot\int_{|k|=\Lambda}\,u_{\ell}^{(0)}\,G_{\ell}(k)\\
 & + & 2\cdot4\cdot\frac{\Lambda^{d}}{(2\pi)^{d}}\cdot\int_{|k|=\Lambda}\,\big(u_{\ell}^{(0)}+u_{\ell}^{(1)}\,||k||+u_{\ell}^{(2)}\,k^{2}\big)\,G_{\ell}(k)\\
 & = & 2\cdot\frac{S_{d}\,\Lambda^{d}}{(2\pi)^{d}}\cdot\frac{6\,u_{\ell}^{(0)}+4\,u_{\ell}^{(1)}\,\Lambda+4\,u_{\ell}^{(2)}\,\Lambda^{2}}{r_{\ell}^{(0)}+r_{\ell}^{(2)}\Lambda^{2}}.
\end{eqnarray*}
The flow of the momentum-dependent term of the self-energy then follows
from \prettyref{eq:correction_self_energy_infinitesimal_anomalous-1},
\prettyref{eq:q_squared_S4}, and \prettyref{eq:projection_self_energy}
as
\begin{eqnarray*}
\ell\frac{d}{d\ell}r_{\ell}^{(2)} & = & 2\cdot4\cdot\frac{\Lambda^{d}}{(2\pi)^{d}}\cdot\int_{|k|=\Lambda}\,\big(\frac{u_{\ell}^{(1)}}{2\Lambda}\,\Big(1-\frac{1}{d}\Big)+u_{\ell}^{(2)}\big)\,G_{\ell}(k).\\
 & = & 8\cdot\frac{S_{d}\,\Lambda^{d}}{(2\pi)^{d}}\cdot\frac{\frac{u_{\ell}^{(1)}}{2\Lambda}\,\Big(1-\frac{1}{d}\Big)+u_{\ell}^{(2)}}{r_{\ell}^{(0)}+r_{\ell}^{(2)}\Lambda^{2}}.
\end{eqnarray*}

\subsection{Detailed calculation of fixed points and critical exponents for $3\le d\le4$\label{sec:Detailed-calculation-of-FP}}

We may now compute the values of the fixed point for arbitrary dimensions
between $3$ and $4$. We again determine $\eta$ such that $r_{\ast}^{(2)}\equiv1$. 

\subsubsection*{Interaction $u$}

The fixed point for the momentum-independent four point coupling must
obey $0\stackrel{!}{=}du_{\ell}^{(0)}/d\ell$ from which follows with
\prettyref{eq:flow_eq_u0}

\begin{eqnarray}
0 & = & (4-d-2\eta)\,u_{\ast}^{(0)}-36\cdot\frac{S_{d}\,\Lambda^{d}}{(2\pi)^{d}}\cdot\frac{(u_{\ast}^{(0)})^{2}}{(r_{\ell}^{(0)}+\Lambda^{2})^{2}}\nonumber \\
 & \stackrel{r_{\ast}^{(0)}\ll\Lambda^{2}}{\simeq} & (4-d-2\eta)\,u_{\ast}^{(0)}-36\cdot\frac{S_{d}\,\Lambda^{d-4}}{(2\pi)^{d}}\cdot(u_{\ast}^{(0)})^{2}\nonumber \\
\Lambda^{d-4}\,u_{\ast}^{(0)} & \simeq & (2\pi)^{d}\,\frac{4-d-2\eta}{36\,S_{d}}.\label{eq:u_0_fixed-1}
\end{eqnarray}
The first-order momentum dependence becomes with $0\stackrel{!}{=}du_{\ell}^{(1)}/d\ell$
and \prettyref{eq:flow_eq_u1}
\begin{eqnarray}
0 & = & (3-d-2\eta)\,u_{\ast}^{(1)}-36\,\Big(2(3-d)-3\eta\Big)\,\frac{S_{d-1}}{d-1}\,\frac{\Lambda^{d}}{(2\pi)^{d}}\,\frac{(u_{\ast}^{(0)})^{2}}{(r_{\ell}^{(0)}+\Lambda^{2})^{2}}\,\frac{1}{\Lambda}\nonumber \\
 & \stackrel{r_{\ast}^{(0)}\ll\Lambda^{2}}{\simeq} & (3-d-2\eta)\,u_{\ast}^{(1)}-36\,\Big(2(3-d)-3\eta\Big)\,\frac{S_{d-1}}{d-1}\,\frac{\Lambda^{d-5}}{(2\pi)^{d}}\,(u_{\ast}^{(0)})^{2}\nonumber \\
u_{\ast}^{(1)} & \simeq & 36\,\frac{2(3-d)-3\eta}{3-d-2\eta}\,\frac{S_{d-1}}{d-1}\,\frac{\Lambda^{d-5}}{(2\pi)^{d}}\,(u_{\ast}^{(0)})^{2}\nonumber \\
\Lambda^{d-3}u_{\ast}^{(1)} & \simeq & \frac{(2\pi)^{d}}{36}\,\frac{2(3-d)-3\eta}{3-d-2\eta}\,(4-d-2\eta)^{2}\frac{S_{d-1}}{(d-1)S_{d}^{2}}.\label{eq:u_1_fixed-1}
\end{eqnarray}
The quadratic momentum dependence of the interaction with $0\stackrel{!}{=}du_{\ell}^{(2)}/d\ell$
and \prettyref{eq:flow_eq_u2}
\begin{eqnarray}
0 & = & (2-d-2\eta)\,u_{\ast}^{(2)}-36\,\Big(2(3-d)-3\eta\Big)\,\frac{4-d-3\eta}{4d}\,\frac{S_{d}\,\Lambda^{d}}{(2\pi)^{d}}\,\frac{(u_{\ast}^{(0)})^{2}}{(r_{\ast}^{(0)}+\Lambda^{2})^{2}}\,\frac{1}{\Lambda^{2}}\nonumber \\
 & \stackrel{r_{\ast}^{(0)}\ll\Lambda^{2}}{\simeq} & (2-d-2\eta)\,u_{\ast}^{(2)}-36\,\Big(2(3-d)-3\eta\Big)\,\frac{4-d-3\eta}{4d}\,\frac{S_{d}\,\Lambda^{d-6}}{(2\pi)^{d}}\,(u_{\ast}^{(0)})^{2}\nonumber \\
u_{\ast}^{(2)} & \simeq & 36\,\frac{2(3-d)-3\eta}{2-d-2\eta}\,\frac{4-d-3\eta}{4d}\,\frac{S_{d}\,\Lambda^{d-6}}{(2\pi)^{d}}\,(u_{\ast}^{(0)})^{2}\nonumber \\
\Lambda^{d-2}u_{\ast}^{(2)} & \simeq & \frac{(2\pi)^{d}}{36\,S_{d}}\,\frac{2(3-d)-3\eta}{2-d-2\eta}\,\frac{(4-d-3\eta)(4-d-2\eta)^{2}}{4d}.\label{eq:u_2_fixed-1}
\end{eqnarray}

\subsubsection*{Mass $r$}

To value for $r_{\ast}^{(0)}$, in the approximation $r_{\ast}^{(0)}\ll\Lambda^{2}$,
is determined from $0\stackrel{!}{=}dr_{\ell}^{(0)}/d\ell$ and \prettyref{eq:flow_eq_r0}

\begin{eqnarray}
0 & = & (2-\eta)\,r_{\ast}^{(0)}+2\cdot\frac{S_{d}\,\Lambda^{d}}{(2\pi)^{d}}\cdot\frac{6\,u_{\ast}^{(0)}+4\,u_{\ast}^{(1)}\,\Lambda+4\,u_{\ast}^{(2)}\,\Lambda^{2}}{r_{\ast}^{(0)}+\Lambda^{2}}\nonumber \\
 & \stackrel{r_{\ast}^{(0)}\ll\Lambda^{2},\eta\ll2}{\simeq} & 2\,r_{\ast}^{(0)}+2\cdot\Lambda^{2}\,\frac{S_{d}\,\Lambda^{d-4}}{(2\pi)^{d}}\cdot\Big(6\,u_{\ast}^{(0)}+4\,u_{\ast}^{(1)}\,\Lambda+4\,u_{\ast}^{(2)}\,\Lambda^{2}\Big)\nonumber \\
\Lambda^{-2}r_{\ast}^{(0)} & \simeq & -\frac{S_{d}}{(2\pi)^{d}}\cdot\Big(6\,\Lambda^{d-4}u_{\ast}^{(0)}+4\,\Lambda^{d-3}u_{\ast}^{(1)}+4\,\Lambda^{d-2}u_{\ast}^{(2)}\Big)\nonumber \\
 & \simeq & -\frac{6\,S_{d}}{(2\pi)^{d}}\cdot\,\Lambda^{d-4}u_{\ast}^{(0)}\label{eq:r0_of_u0-1}\\
 &  & -\frac{144\,S_{d}}{(2\pi)^{2d}}\,\frac{2(3-d)-3\eta}{3-d-2\eta}\,\Big(\frac{S_{d-1}}{d-1}+\frac{4-d-3\eta}{4d}\,S_{d}\Big)\,(\Lambda^{d-4}u_{\ast}^{(0)})^{2}.\nonumber 
\end{eqnarray}

\paragraph*{Critical exponent $\eta$}

Demanding $0\stackrel{!}{=}dr_{\ell}^{(2)}/d\ell$ the anomalous dimension
$\eta$ is determined from \prettyref{eq:flow_eq_r2} as 
\begin{eqnarray}
\eta & = & 8\,\frac{S_{d}\,\Lambda^{d}}{(2\pi)^{d}}\cdot\frac{\frac{u_{\ast}^{(1)}}{2\Lambda}\,\Big(1-\frac{1}{d}\Big)+u_{\ast}^{(2)}}{r_{\ast}^{(0)}+\Lambda^{2}}\nonumber \\
 & \stackrel{r_{\ast}^{(0)}\ll\Lambda^{2}}{\simeq} & 8\,\frac{S_{d}\,\Lambda^{d-2}}{(2\pi)^{d}}\cdot\Big(\frac{u_{\ast}^{(1)}}{2\Lambda}\,\Big(1-\frac{1}{d}\Big)+u_{\ast}^{(2)}\Big).\label{eq:approx_eta_d34-1}
\end{eqnarray}

Inserting \prettyref{eq:u_0_fixed}, \prettyref{eq:rel_u1}, and \prettyref{eq:rel_u2},
we get
\begin{eqnarray*}
\eta & \simeq & 8\,\frac{S_{d}}{(2\pi)^{d}}\cdot\Big(\frac{1}{2}\Big(1-\frac{1}{d}\Big)\,\Lambda^{d-3}u_{\ast}^{(1)}+\Lambda^{d-2}u_{\ast}^{(2)}\Big)\\
 & = & 8\,\frac{S_{d}}{(2\pi)^{2d}}\cdot\Big(\frac{1}{2}\Big(1-\frac{1}{d}\Big)\,36\,\frac{2(3-d)-3\eta}{3-d-2\eta}\,\frac{S_{d-1}}{d-1}\\
 & + & 36\,\frac{2(3-d)-3\eta}{2-d-2\eta}\,\frac{4-d-3\eta}{4d}\,S_{d}\Big)\,\Lambda^{2d-8}\,(u_{\ast}^{(0)})^{2}.
\end{eqnarray*}

\paragraph*{Critical exponent $\nu$}

To determine the critical exponent $\nu$, we need to linearize the
flow about the fixed point and determine the eigenvalue $\lambda_{r}$
in the direction of the parameter $r$. We have the general form of
the flow equation
\begin{eqnarray*}
\ell\,\frac{d}{d\ell}\left(\begin{array}{c}
r_{\ell}^{(0)}\\
r_{\ell}^{(2)}\\
u_{\ell}^{(1)}\\
u_{\ell}^{(1)}\\
u_{\ell}^{(2)}
\end{array}\right) & = & \beta(r_{\ell}^{(0)},r_{\ell}^{(2)},u_{\ell}^{(0)},u_{\ell}^{(1)},u_{\ell}^{(2)}).
\end{eqnarray*}
Since the influence of $r_{\ell}^{(0)}$ on the flow equations of
the $u_{\ell}^{(n)}$ is negligible, we can also neglect all terms
$\frac{\partial\beta_{u^{(n)}}}{\partial r_{\ell}^{(0)}}\simeq0$.
Linearizing the flow equation \prettyref{eq:flow_eq_r0} with $\delta r_{\ell}^{(0)}=r_{\ell}^{(0)}-r_{\ast}^{(0)}$,
we get

\begin{eqnarray*}
\ell\frac{d\delta r_{\ell}^{(0)}}{d\ell} & = & (2-\eta)\,\delta r_{\ell}^{(0)}-2\cdot\frac{S_{d}\,\Lambda^{d}}{(2\pi)^{d}}\cdot\frac{6\,u_{\ast}^{(0)}+4\,u_{\ast}^{(1)}\,\Lambda+4\,u_{\ast}^{(2)}\,\Lambda^{2}}{(r_{\ast}^{(0)}+\Lambda^{2})^{2}}\,\delta r_{\ell}\\
 & \simeq & (2-\eta)\,\delta r_{\ell}^{(0)}-2\cdot\frac{S_{d}}{(2\pi)^{d}}\cdot\big(6\,\Lambda^{d-4}u_{\ast}^{(0)}+4\,\Lambda^{d-3}u_{\ast}^{(1)}+4\,\Lambda^{d-2}u_{\ast}^{(2)}\big)\,\delta r_{\ell}\\
 & \stackrel{(\ref{eq:u_0_fixed-1}),(\ref{eq:u_1_fixed-1}),(\ref{eq:u_2_fixed-1})}{\simeq} & (2-\eta)\,\delta r_{\ell}^{(0)}\\
 &  & -\frac{1}{18}\,(4-d-2\eta)(2(3-d)-3\eta)\,\big(\frac{6}{2(3-d)-3\eta}+\frac{4}{d-1}\,\frac{4-d-2\eta}{3-d-2\eta}\,\frac{S_{d-1}}{S_{d}}\\
 &  & +\frac{1}{d}\,\frac{(4-d-3\eta)(4-d-2\eta)}{2-d-2\eta}\big)\,\delta r_{\ell}\\
 & = & \lambda_{r}\,\delta r.
\end{eqnarray*}
The critical exponent $\nu=\lambda_{r}^{-1}$ thus takes the form
\prettyref{eq:nu_general}, as stated in the main text.

\section{Bibliography}

\providecommand{\newblock}{}

\end{document}